\documentclass[conference]{IEEEtran}
\IEEEoverridecommandlockouts
\usepackage{color}
\usepackage{graphicx}  % Written by David Carlisle and Sebastian Rahtz
\usepackage{psfrag}    % Written by Craig Barratt, Michael C. Grant,
\usepackage{algorithm}
\usepackage{algorithmic}
\usepackage{amsfonts,amssymb}
\usepackage{amsthm}
\usepackage{amsmath}
\usepackage{amstext}
\usepackage{bm}
\usepackage{subfigure}
\usepackage{hyperref}
\usepackage{CJK}
\usepackage{url}
\usepackage{diagbox}
\usepackage[square, comma, sort&compress, numbers]{natbib}
\usepackage{cases}
\begin{document}

\title{Channel-Adaptive Robust Resource Allocation for Highly Reliable IRS-Assisted V2X Communications}
\author{Peng Wang, Weihua Wu 
%Jiayi Liu, Guanhua Chai and Li Feng
%\thanks{Copyright (c) 2015 IEEE. Personal use of this material is permitted. However, permission to use this material for any other purposes must be obtained from the IEEE by sending a request to pubs-permissions@ieee.org.}
\thanks{*This work was supported in part by the NSF China under Grant 61801365; in part by the Natural Science Foundation of Shaanxi Province under Grant 2023-JC-YB-542; in part by the China Postdoctoral Science Foundation under Grant 2018M643581; in part by Jiangsu Province Postdoctoral Foundation under Grant 2020Z208; and in part by Natural Science Foundation of Jiangsu Province under Grant BK20200904. (Corresponding author: Weihua Wu.)}
\thanks{P. Wang, is with State Key Laboratory of ISN, School of Telecommunications Engineering, Xidian University, No.2 South Taibai Road, Xi'an, 710071, Shaanxi, China (email: wangp134@163.com).}
\thanks{W. Wu is with the School of Physics and Information Technology, Shaanxi Normal University, Xi'an 710119, China (email: whwu@snnu.edu.cn).}
%\thanks{L. Feng is with School of Computer Science and Communication Engineering, Jiangsu University, No. 301 Xuefulu, Zhenjiang, 212013, Jiangsu, China (email: yunxuan@ujs.edu.cn).}

}
\maketitle

%while guaranteeing the minimum signal-to-noise ratio (SINR) requirement of certain throughput for V2I communications and the probabilistic quality-of-service (QoS) constraints for V2V communications

\begin{abstract}
This paper addresses the challenges of resource allocation in vehicular networks enhanced by Intelligent Reflecting Surfaces (IRS), considering the uncertain Channel State Information (CSI) typical of vehicular environments due to the Doppler shift. 
Leveraging the 3GPP's Mode 1 cellular V2X architecture, our system model facilitates efficient subcarrier usage and interference reduction through coordinated V2I and V2V communications. 
Each Cellular User Equipment (CUE) shares its spectrum with at most one Vehicular User Equipment (VUE) in a one-to-one reuse pattern.
We formulate a joint optimization problem for vehicular transmit power, Multi-User Detection (MUD) matrices, V2V link spectrum reuse, and IRS reflection coefficients in IRS-aided V2X communication with imperfect CSI. 
To tackle this, a novel robust resource allocation algorithm is developed by first decomposing the problem into manageable sub-problems such as power allocation, MUD matrices optimization and IRS phase shifts, and then using the Block Coordinate Descent (BCD) method to alternately optimize these subproblems for optimal resource allocation.
Our contributions include efficient approaches for self-learning based power allocation and phase shift optimization that adapt to CSI uncertainties, significantly enhancing the reliability and efficiency of vehicular communications. Simulation results validate the effectiveness of the proposed solutions in improving the Quality of Service (QoS) and managing the complex interference inherent in dense vehicular networks.
%This paper addresses the challenges of resource allocation in vehicular networks enhanced by Intelligent Reflecting Surfaces (IRS), considering the uncertain Channel State Information (CSI) typical of vehicular environments. 
%Leveraging the 3GPP's Mode 1 cellular V2X architecture, our system model facilitates efficient subcarrier usage and interference reduction through coordinated V2I and V2V communications, where each Cellular User Equipment (CUE) can share spectrum with at most one Vehicular User Equipment (VUE). 
%Then, we  formulate a joint vehicular transmit power, Multi-User Detection (MUD) matrices, V2V link spectrum reuse, and IRS reflection coefficients optimization problem for IRS-aided V2X communication with imperfect CSI. Specifically, the total capacity of all cellular user equipments (CUEs) is maximized adhere to the outage probability constraints of vehicular user equipments.
%To tackle this, a robust resource allocation algorithm is introduced to decompose the problem into manageable sub-problems such as power allocation and IRS phase shifts using the Block Coordinate Descent (BCD) method. Our contributions include novel approaches for self-learning power allocation and phase shift optimization that adapt to the CSI uncertainties, significantly enhancing the reliability and efficiency of vehicular communications. Simulation results validate the effectiveness of the proposed solutions in improving the Quality of Service (QoS) and managing the complex interference inherent in dense vehicular networks.

\vspace{5pt}
\textbf{\emph{Key Terms}}: V2X communications, IRS, robust resource allocation, outage probability constraint, uncertain channel.

\end{abstract}


%for peerreview papers, inserts a page break and creates the second title.
% Will be ignored for other modes.

\section{Introduction}

\newtheorem {theorem}{\textbf{Theorem}}
\newtheorem {lemma}{\textbf{Lemma}}
\newtheorem {remark}{\textbf{Remark}}
\newtheorem {definition}{\textbf{Definition}}

In recent years, the integration of ground and aerial vehicles within Intelligent Transportation Systems (ITS) has advanced, driven by companies like BMW and Tesla developing autonomous driving and assistance systems \cite{shah20185g}. ITS depends on reliable vehicular communication, leading to the evolution of Vehicle-to-Everything (V2X) communications, including Vehicle-to-Infrastructure (V2I) and Vehicle-to-Vehicle (V2V) \cite{ganesan2020nr,guo2019resource}. Key technologies include Dedicated Short-Range Communications (DSRC) and cellular-based vehicular networks \cite{naik2019ieee}. V2X supports traffic management, road safety, autonomous driving, and in-vehicle entertainment, but faces performance and security challenges. 
%Key technologies include Dedicated Short-Range Communications (DSRC) and cellular-based vehicular networks \cite{naik2019ieee}. DSRC has limitations in coverage, data rates, Quality of Service (QoS), and latency in high-density, high-mobility environments. To address these, the 3rd Generation Partnership Project (3GPP) is developing Cellular Vehicle-to-Everything (C-V2X) \cite{etsi2013intelligent}, which must also tackle resource allocation challenges.

In V2X communications, the high mobility of vehicles reduces the stability of channels due to the doppler effect, thereby decreasing transmission rates. To improve the Quality of Service (QoS) of V2X communications and enhance transmission efficiency, Intelligent Reflecting Surfaces (IRS) have emerged as a burgeoning technology attracting increasing attention \cite{9733238}. IRS is a passive array structure that can modify the signal propagation path by continuously or discretely adjusting the phase of each reflective unit on the surface, with extremely low power consumption \cite{9722826}. Specifically, IRS aims to reflect the base station's signals to specific receivers, enhancing the received signal power or suppressing interference to ensure communication security or privacy \cite{yu2020robust}. IRS can be easily deployed on urban infrastructure, such as building facades or roadside facilities, which reduces the operators' costs and the complexity of installations \cite{renzo2019smart}. Furthermore, IRS can effectively guide signals to areas that are typically hard to cover, expanding the coverage of V2X communications. When the Line of Sight (LoS) link between the sender and receiver in vehicular networks is blocked by obstacles, IRS can create new propagation paths through reflection, avoiding penetration losses and maintaining optimal transmission rates \cite{9424177}. Given these advantages, resource scheduling in IRS-assisted wireless communication systems has become a hot research topic.

Although IRSs are effective in optimizing channel performance and dynamically adjusting reflection properties according to communication needs, they also face some challenges. 
One of the most significant challenges is accurately acquiring Channel State Information (CSI) in the highly dynamic environment of vehicular networks \cite{coll2019sub, wang2023joint}. Due to the high-speed movement of vehicles, channel characteristics may undergo rapid and frequent changes, making it extremely difficult to accurately capture these changes in real time \cite{wu2021learning, liang2019spectrum}. 
Moreover, it is difficult to manage the each phase shift of IRS in the above complex and uncertain channel environment \cite{9826440}.
To fully leverage the advantages of IRS, it is necessary to develop efficient algorithms that can dynamically adjust the phases of IRS units to optimize the reflection path of signals, considering the unique challenges of the vehicular environment, such as the high-speed movement and instability of signals. These algorithms must be capable of real-time adaptation to the changing channel conditions to maintain optimal communication performance \cite{9497709}.
By implementing well-designed resource allocation strategies, it is possible to mitigate the effects of uncertain CSI and effectively allocate all kinds of resource in IRS-aided V2X communications.

The study investigated the resource allocation problem in IRS-assisted vehicular networks under the condition of uncertain CSI for vehicular channels.
In practical vehicular network environments, the Next Generation NodeB (gNB) can typically obtain accurate CSI for V2I links, whereas CSI estimation for V2V links inevitably suffers from inaccuracies due to inherent channel state uncertainties.
This work is based on the Mode 1 defined in the 3GPP cellular V2X architecture, where gNB can collect comprehensive vehicle information, enabling more efficient subcarrier utilization and reducing interference. 
To address the issue of limited spectrum, we propose a system model where V2I communications are assigned radio spectrum orthogonally, while V2V communications reuse the radio spectrum allocated to V2I communications. 
Each Cellular User Equipment (CUE) can share its spectrum with at most one Vehicular User Equipment (VUE), and each VUE can at most reuse the spectrum of one CUE. 
This design is necessary, but it leads to a more complex interference management scheme. 
Subsequently, a joint optimization problem of vehicular transmission power, Multi-User Detection (MUD) matrix, V2V link spectrum reuse, and IRS reflection coefficients is designed to maximize the total capacity of all CUEs, subject to the interruption probability QoS constraints of the VUEs. 
Due to the non-convexity of outage probability constraints, constant modulus constraints on IRS phases and the objective function, this joint resource allocation problem is a complex non-convex optimization problem.
To solve this resource allocation problem, the proposed optimization problem is decomposed into four sub-problems using the Block Coordinate Descent (BCD) method, with each sub-problem solved independently. 
However, all three sub-problems are still non-convex due to the presence of outage probability constraints for uncertain channels and constant modulus constraints on IRS phases. Therefore, this paper designs appropriate resource allocation methods for different sub-problems.
The contributions of this paper are summarized as follows:

\begin{itemize}
  \item For the power allocation sub-problem, a self-learning power allocation algorithm is proposed, which includes a learning method and a graph-based analytic mapping approach. In this algorithm, channel uncertainties are learned through an affine set, transforming the intractable chance constraints into simple linear constraints. Then, a graph-based analytic mapping approach is utilized to derive closed-form solutions for the non-convex optimization problem.
  \item To optimize the IRS phase shift sub-problem, a phase shift allocation algorithm is developed, which contains a Conditional Value-at-Risk (CVaR) approximation method, a Continuous Convex Approximation (CCA) method and a Semidefinite Relaxation (SDR) approach. First, CCA approach and SDR approach are employed to transform the non-convex objective function and constant modulus constraint into the tractable form. Then, CVaR approximation method is used to utilize uncertain channel state information, equivalently transforming the chance constraints into linear matrix inequality constraints, resulting in a low-complexity optimization approach.
  \item Based on all these distinct sub-problems, a adaptive channel-based alternating optimization (ACAO) algorithm is designed to iteratively optimize the MUD matrix, power allocation, and IRS phase shift optimization sub-problems alternately based on the given spectrum reuse pattern, achieving optimal resource allocation. Then, the optimal spectrum reuse is determined for all possible spectrum reuse pairs.
\end{itemize}

The remainder of this paper is organized as follows. In Section II, we review related work. Section III describes the system model and formulates the optimization problem. Section IV proposes a robust resource allocation algorithm to solve the optimization problem and analyzes the algorithm's complexity and convergence. Finally, simulation results are discussed in Section V, and the paper is concluded in Section VI.

\section{Related Work}

In recent years, IRS has demonstrated substantial potential in enhancing the performance of wireless communication systems, notably in improving signal coverage, spectral efficiency, and energy efficiency. Within this context, significant research has been undertaken on IRS-assisted wireless communication systems, as evidenced by \cite{li2020reconfigurabledvxvd,zhou2020intelligentvcvxrr,pan2020multicell,basar2021indoor}. While research on IRS-assisted wireless networks is well-established, studies on IRS-assisted vehicular networks are still in their early stages. As computational demands in autonomous driving applications grow, the importance of this field is expected to increase. Current research in IRS-assisted vehicular network communications predominantly employs a perfect CSI model, overlooking the mobility of vehicular users. The study in \cite{chen2020resource} explores the resource allocation in IRS-assisted vehicular communications based on slow-varying large-scale fading channel information, where power allocation, IRS reflection coefficients, and spectrum allocation are jointly optimized. To accommodate various QoS requirements in vehicular communications, this research maximizes the overall capacity of V2I links while ensuring a minimum Signal-to-Interference-plus-Noise Ratio (SINR) for V2V links. The work in \cite{jia2020reconfigurable} investigates joint power control and passive beamforming for D2D users in IRS-assisted D2D communication networks, aimed at maximizing energy efficiency. To address the challenges of rapidly changing V2X channels and the difficulty of acquiring instantaneous CSI, \cite{chen2021qos} studies the spectrum sharing in IRS-assisted vehicular networks based on large-scale slow fading channel information, where V2V links reuse the spectrum allocated to V2I links to maximize the total capacity of V2I links while ensuring the reliability of V2V links used for exchanging safety information.

Most of the aforementioned methods focus solely on the large-scale fading of vehicular links, insufficiently considering channel uncertainties caused by factors such as the Doppler effect. In real vehicular communication scenarios, the dynamic link state induced by high-speed moving vehicles can complicate the estimation of CSI. These works are unable to guarantee interruption probability constraints, potentially leading to violations of communication link QoS constraints. To uphold QoS constraints, \cite{cheng2022robust} investigates the joint beamforming issue under both perfect and imperfect CSI conditions and proposes a method based on SDR and BCD. For different CSI scenarios, this study develops efficient algorithms including analyses of convergence and complexity. \cite{zhou2020framework} explores robust beamforming algorithms based on imperfect cascaded channel information at the transmitter end. It formulates problems of minimizing transmission power under rate constraints for the worst-case scenario with bounded CSI errors and under rate interruption probability constraints based on statistical CSI error models. \cite{al2022reconfigurable} examines user scheduling in high-mobility scenarios within IRS-assisted systems using deep reinforcement learning techniques, which can significantly compensate for channel gain losses caused by high vehicular mobility and substantially increase the total capacity of vehicular communications. \cite{chen2021robust} investigates a robust transmission scheme for time-varying IRS-assisted millimeter-wave vehicular communications. Considering imperfect statistical CSI, an effective transmission protocol is proposed, followed by studies on resource allocation issues under single and multiple VUE scenarios. However, these studies generally assume that CSI is based on known probabilistic distribution models, such as statistical error models and bounded error models, rather than on unknown probabilistic distribution models.

\section{System Model}

%\begin{figure*}
%\begin{center}
%\includegraphics[width=6.6in,height=2.8in]{system_model.eps}
%\label{fig2}\caption{System model.}
%\end{center}
%\end{figure*}
\subsection{Network Model}

Our system model, as depicted in Fig. 1, considers an IRS-aided vehicular communication system in a single-input multiple-output (SIMO) uplink scenario, where there exists one multi-antenna gNB, one IRS and multiple single-antenna vehicles.
We assume that gNB is equipped with $M$ antennas and IRS contains $N$ programmable phase shifters.
The vehicles in the system model are divided into $I$ CUEs that communicate with the gNB via V2I links, and $L$ VUE pairs that perform V2V communications.
The set of CUEs and VUE pairs is defined as ${\cal I} = \{ 1,2, \cdots ,I\} $ and ${\cal L} = \{ 1,2, \cdots ,L\} $, respectively.
In mode-1 of NR-V2X, the orthogonal resource blocks (RBs) are allocated to CUEs by the gNB \cite{etsi2013intelligent}.
As the uplink resources are less intensively utilized by CUEs and VUE pairs produce small interference to the gNB, VUE pairs reuse the uplink resources of CUEs.
We assume that one VUE pair can only reuse the spectrum of at most one CUE and the spectrum of one CUE can only be reused by at most one VUE pair.
The binary variable ${{x_{i,l}}}$ is applied to represent the allocation of uplink resources for CUE.
If the spectrum of the $i$th CUE is reused by the $l$th VUE, ${{x_{i,l}}}=1$. Otherwise, ${{x_{i,l}}}=0$.
Let IRS reflection coefficient matrix be defined as a $N \times N$ diagonal matrix ${\bf{\Phi}} = diag\left( {\left[ {{e_1}, \cdots ,{e_N}} \right]} \right)$, where $e_n=e^{j\theta_n}$ and $\theta_n \in [0,2\pi]$ is the phase shift of the $n$th reflecting element.
The MUD matrix is employed at the gNB to facilitate multi-user detection by separating and decoding the signals from different CUEs using the spatial diversity of the multiple antennas, denoted as $\mathbf{F}=[\mathbf{f}_1, \cdots, \mathbf{f}_I] \in {\mathbb{C}^{M \times I}}$, where $\mathbf{f}_i$ is the $i$th column of the matrix $\mathbf{F}$.
The transmit power of the $i$th CUE and the transmitter of the $l$th VUE pair is represented as ${P_i^c}$ and ${P_l^v}$, respectively.

\begin{figure}
\setlength{\abovecaptionskip}{-0.5em}
\begin{center}
\includegraphics[scale = 0.25]{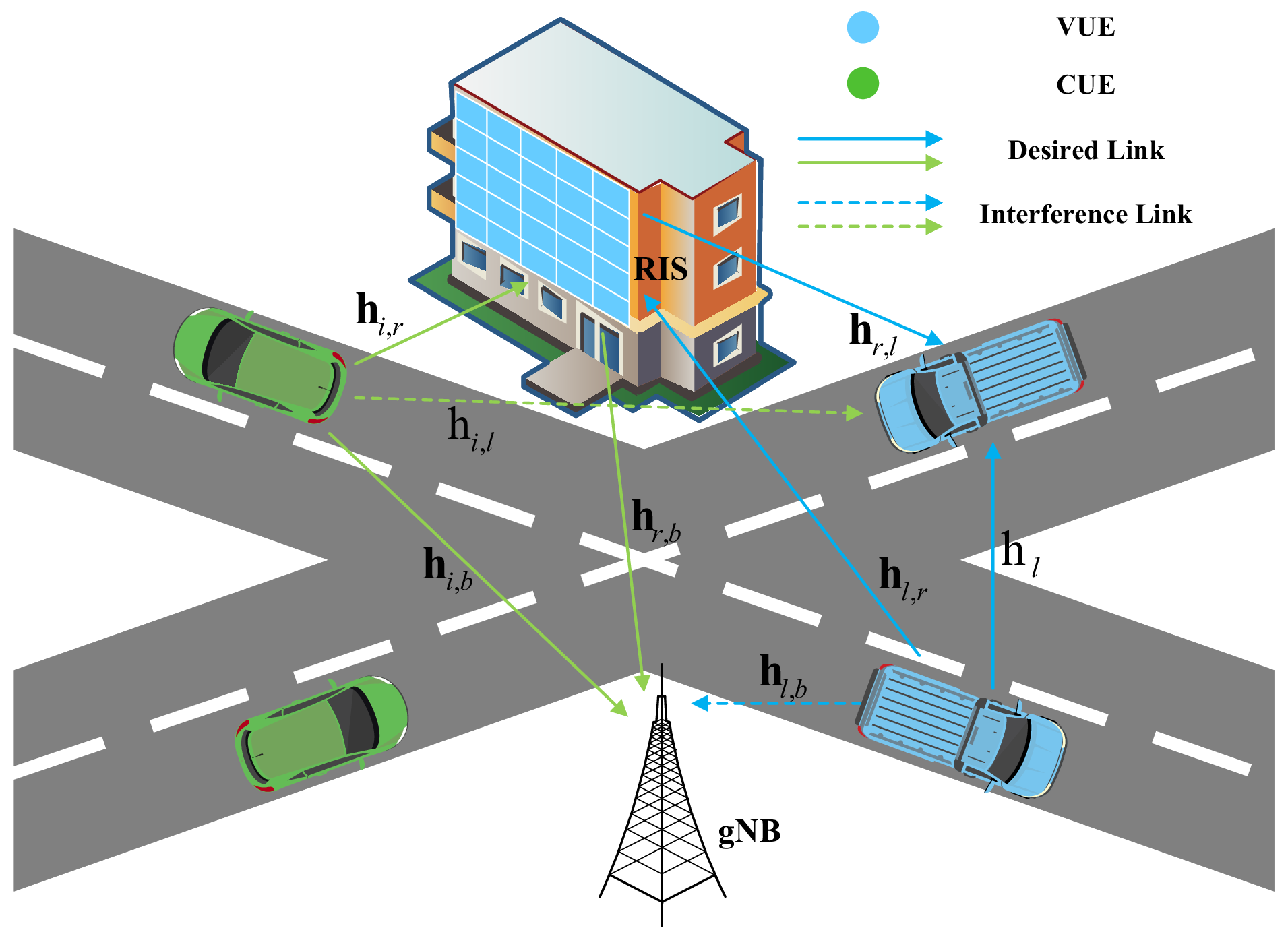}
\caption{IRS-aided single cellular vehicle communications}\label{gdfgdger}
\end{center}
\vspace{-2em}
\end{figure}

The channel gain from the $i$th CUE to the gNB can be modelled as ${{\bf{g}}_{i,b}} = \sqrt {\alpha d_{i,b}^{{\lambda _{i,b}}}} { {\bf{h}}_{i,b}} = \sqrt {\beta_{i,b}} { {\bf{h}}_{i,b}}\in {\mathbb{C}^{M \times 1}}$, where $\beta_{i,b}$ is the large-scale slow fading channel gain, ${ {\bf{h}}_{i,b}}$ represents the small-scale fast fading effect, $\alpha $ is the pathloss at the distance of 1 metre, $d_{i,b}$ is the distance between the gNB and the $i$th CUE in metres and $\lambda_{i,b}$ represents the corresponding pathloss exponent.
Similarly, we suppose that the channel gain from the $i$th CUE to the IRS, the channel gain from the IRS to the gNB, the channel gain from the transmitter of the $l$th VUE pair to the IRS, the channel gain from the IRS to the receiver of the $l$th VUE pair and the channel gain between the $l$th VUE pair are represented as ${{\bf{g}}_{i,r}} \in {\mathbb{C}^{N \times 1}}$, ${{\bf{g}}_{r,b}} \in {\mathbb{C}^{N \times M}}$, ${{\bf{g}}_{l,r}} \in {\mathbb{C}^{N \times 1}}$, ${{\bf{g}}_{r,l}} \in {\mathbb{C}^{N \times 1}}$ and ${g_l}$, respectively.
Moreover, the crosstalk channel gain from $i$th CUE to the receiver of the $l$th VUE pair and the crosstalk channel gain from the transmitter of the $l$th VUE pair to the gNB are denoted as ${g_{i,l}}$ and ${{\bf{g}}_{l,b}}$, respectively.
We assume that the above definitions of the channel gain are all similar to ${{\bf{g}}_{i,b}}$.
Because the large-scale fading components are primarily affected by vehicle position and vary on a slow scale, the large-scale fading components of all links can be accurately obtained by gNB.
However, for the small-scale fading, we consider different assumptions for different channels.
Since gNB has perceptive capabilities, we assume that gNB can obtain the small-scale fading of the channels which directly connected to gNB, i.e. ${{\bf{h}}_{i,b}}$ and ${{\bf{h}}_{l,b}}$, and the cascaded channels whose final destination is gNB, i.e. ${{{\bf{h}}}_{i,r}}$, ${{{\bf{h}}}_{l,r}}$ and ${{{\bf{h}}}_{r,b}}$.
For the channels whose destinations are vehicles, i.e. ${ h_l}$, ${ h_{i,l}}$ and ${{{\bf{h}}}_{r,l}}$, we assume that the gNB can only obtain the estimated channel fading $\bar h$ with error $\hat h$, since the high mobility of vehicles will generate the Doppler shift.
The first-order Gauss-Markov process \cite{kim2010doesgdgcb} is applied to model the small-scale fading of the vehicular channels as
\begin{align}
 h = \tau \bar h + \sqrt {1 - {\tau ^2}} \hat h,
\end{align}
where $\hat h \sim {\cal C}{\cal N}(0,1)$ is independent and identically distributed (i.i.d.).
The coefficient $\tau \;\left( {0 < \tau  < 1} \right)$ quantifies the channel correlated at two successive time slots and is given as
\begin{align}
\tau=J_0 (2\pi f_s T),\nonumber
\end{align}
where $J_0$ is the zero-order Bessel function of the first kind, $T$ is a period feedback latency.
$f_s=vf_c/c$ is the maximum Doppler frequency, where $f_c$ is the carrier frequency, $v$ is the vehicle speed and $c=3 \times 10^5 km/s$.

Herein, the received signal of the gNB from the $i$th CUE is given by
\begin{align}
y_i = & \mathbf{f}_i^H \left[ \sqrt{P_i^c} \left( \mathbf{g}_{i,b} + \mathbf{g}_{r,b}^H\mathbf{\Phi}\mathbf{g}_{i,r} \right) c_i \right. \nonumber\\
& \left. + \sum_{l \in \mathcal{L}} x_{i,l}\sqrt{P_l^v} \left( \mathbf{g}_{r,b}^H\mathbf{\Phi}\mathbf{g}_{l,r} + \mathbf{g}_{l,b} \right) c_l + \mathbf{n} \right],
\end{align}
where ${\bf{n}}=[n_1,\cdots,n_M]^T$, $n_m$ is the additive Gaussian white noise and subjects to ${\cal N}\left( {0,{\sigma ^2}} \right)$. $c_i$ and $c_l$ denote the transmitted information symbol of the $i$th CUE and the transmitter of the $l$th VUE pair, respectively.
The received signal of the receiver of the $l$th VUE pair is formulated as
\begin{align}
y_l = & \sqrt{P_l^v} \left( g_l + \mathbf{g}_{r,l}^H \mathbf{\Phi} \mathbf{g}_{l,r} \right) c_l \nonumber\\
& + \sum_{i \in \mathcal{I}} x_{i,l} \sqrt{P_i^c} \left( g_{i,l} + \mathbf{g}_{r,l}^H \mathbf{\Phi} \mathbf{g}_{i,r} \right) c_i + n_l,
\end{align}
where $n_l$ is additive Gaussian white noise and follows a normal distribution given by $n_l \sim {\cal N}\left( {0,{\sigma ^2}} \right)$. Thus, the received SINR at the gNB and the receiver of the $l$th VUE pair are formulated as
\begin{align}
%\vspace{-2em}
\gamma _i &= \frac{{P_i^cg_i^B}}{{\sum\limits_{l \in {\cal L}} {{x_{i,l}}P_l^vg_l^B + {\sigma ^2 |{\bf{f}}_i^H|^2}} }} \nonumber\\
&= \frac{{P_i^c{{\left| {{\bf{f}}_i^H\left( {{{\bf{g}}_{i,b}} + {\bf{g}}_{r,b}^H{\bf{\Phi }}{{\bf{g}}_{i,r}}} \right)} \right|}^2}}}{{\sum\limits_{l \in {\cal L}} {{x_{i,l}}P_l^v{{\left| {{\bf{f}}_i^H\left( {{{\bf{g}}_{l,b}} + {\bf{g}}_{r,b}^H{\bf{\Phi }}{{\bf{g}}_{l,r}}} \right)} \right|}^2} + {\sigma ^2 |{\bf{f}}_i^H|^2}} }},
%\vspace{-2em}
\end{align}
and
\begin{align}
{\gamma _l} = \frac{{P_l^v g_l^v}}{{\sum\limits_{i \in {\cal I}} {{x_{i,l}}P_i^c g_i^v}  + {\sigma ^2}}} = \frac{{P_l^v{{\left| {{g_l} + {\bf{g}}_{r,l}^H{\bf{\Phi }}{{\bf{g}}_{l,r}}} \right|}^2}}}{{\sum\limits_{i \in {\cal I}} {{x_{i,l}}P_i^c{{\left| {{g_{i,l}} + {\bf{g}}_{r,l}^H{\bf{\Phi }}{{\bf{g}}_{i,r}}} \right|}^2}}  + {\sigma ^2}}},
\end{align}
respectively.

\subsection{Problem Formulation}

To satisfy the different QoS requirements for different vehicular links, i.e., large-capacity services for V2I links and ultra-reliable requirements for V2V links, the sum capacity of $I$ CUEs should be maximized while guaranteeing the minimum reliability of $L$ VUE pairs.
Therefore, the resource allocation problem is a joint the transmit power ${\bf{P}} = \left\{ {P_i^c,P_l^v,\forall i,l} \right\}$, the spectrum reuse ${\bf{X}} = \{ {x_{i,l}},\forall i,l\} $, the MUD matrix $\bf{F}$ and the IRS phase shift matrix ${\bf{\Phi}}$ optimization problem, which can be given as
\begin{align}
\!\!\!\!\!\!\!\!\!\!\!\!\!\!\!\!\mathop {\max }\limits_{\{ \mathbf{P},\mathbf{\Phi},\mathbf{X},\bf{F}\} } &  \sum\limits_{i =1}^I {W{{\log }_2}(1 + {\gamma _i})} \label{1}\\
\textrm{s.t.}\quad  & \Pr \{ {\gamma _l} \ge {\gamma _{th}}\}  \ge 1 - \delta ,\forall l, \tag{\ref{1}{a}}\label{1a}\\
&\left| {{e_n}} \right| = 1,\;\forall n,\tag{\ref{1}{b}}\label{1b}\\
&\sum\limits_{l =1}^L {{x_{i,l}}}  = 1,\forall i,\tag{\ref{1}{c}}\label{1c}\\
&\sum\limits_{i =1}^I {{x_{i,l}} = 1,} \forall l,\tag{\ref{1}{d}}\label{1d}\\
&{x_{i,l}} \in \{ 0,1\} ,\forall i,l,\tag{\ref{1}{e}}\label{fdfgre}\\
&0 \!\le\! P_i^c\! \le\! P_{\max }^c,\forall i, \tag{\ref{1}{f}}\label{1f}\\
&0\! \le\! P_l^v\! \le\! P_{\max }^v,\;\forall l,\tag{\ref{1}{g}}\label{1g}
%&\sum\limits_{l \in {\cal L}} {{z_{i,l}}}  = 1,\forall i \in {\cal I},\tag{\ref{1}{c}}\label{1d}\\
%&\sum\limits_{i \in {\cal I}} {{z_{i,l}} = 1,} \forall l \in {\cal L},\nonumber \\
%&0 \le P_i^c \le P_{\max }^c,\forall i \in {\cal I},\tag{\ref{1}{d}}\label{1e}\\
%&0 \le P_l^d \le P_{\max }^v,\;\forall l \in {\cal L},\nonumber
\end{align}
where $W$ is the bandwidth of spectrum resources, ${\gamma _{th}}$ is the SINR threshold to establish the reliable V2V links, $\delta$ is the maximum acceptable outage probability, $P_{\max }^c$ and $P_{\max }^v$ are the maximum transmit powers of CUEs and the transmitters of VUE pairs, respectively.
Specifically, constraint (\ref{1a}) is employed to ensure the outage probability of VUE pairs.
Constraint (\ref{1b}) guarantees the unit-modulus requirements of the reflecting elements at the IRS.
Constraints (\ref{1c}) and (\ref{1d}) indicate that the spectrum of each CUE can only be shared with at most one VUE and each VUE can only reuse the spectrum of at most one CUE, respectively.
Constraints (\ref{1f}) and (\ref{1g}) guarantee that the transmit power of each CUE and VUE transmitter is limited to the maximum transmit power, respectively.

In conclusion, problem (\ref{1}) is a mixed-integer non-convex problem, which is challenging to optimize due to the following three reasons.
First, the variables are highly coupled with each other in the objective function and constraint (\ref{1a}).
Second, the closed-form expression of constraint (\ref{1a}) is difficult to obtain due to its probability form.
Third, the binary variable and the unit-modulus constraint aggravate the difficulty of solving problem (\ref{1}).
Therefore, we propose a robust resource allocation approach to solve this problem in Section IV.

\section{Adaptive Channel-Based Alternating Optimization Algorithm}

In this section, the adaptive channel-based alternating optimization (ACAO) algorithm is presented to solve the joint power ${\bf{P}}$, MUD matrix ${\bf{F}}$, IRS phase shifts ${\bf{\Phi}}$ and spectrum ${\bf{X}}$ robust optimization problem in (\ref{1}). Specifically, problem (\ref{1}) is first decomposed into four subproblems for handling the coupling of the variables.
Then, the power allocation subproblem, the MUD matrix optimization subproblem and the IRS phase shifts optimization subproblem are iteratively optimized in an alternating manner to maximize the capacity of all CUEs when the other variables are fixed based on each possible spectrum reusing pair $\{x_{i,l},\forall i,l\}$.
While fixing the spectrum reusing pair ${\bf{X}}$, the alternating optimization problem for one CUE-VUE pair is reformulated as
\begin{align}
\mathop {\max }\limits_{ \{{\bf{P}}, {\bf{\Phi }}, {\bf{F}} \}} \quad  & {C_{i,l}}={W{{\log }_2}(1 + \frac{{P_i^cg_i^B}}{{P_l^vg_l^B + {\sigma ^2 |{\bf{f}}_i^H|^2}}})} \label{dfsg}\\
\textrm{s.t.}\quad  & \Pr \{ \frac{{P_l^vg_l^v}}{{P_i^cg_i^v + {\sigma ^2}}} \ge {\gamma _{th}}\}  \ge 1 - \delta ,\tag{\ref{dfsg}{a}}\label{dsfgsda}\\
&\left| {{e_r}} \right| = 1,\;\forall n,\tag{\ref{dfsg}{b}}\label{dsfgsdb}\\
& 0 \le P_i^c \le P_{\max }^c,\;\;0 \le P_l^d \le P_{\max }^v.\tag{\ref{dfsg}{c}}\label{dsfgsdc}
\end{align}
Finally, the spectrum subproblem can be solved by the Hungarian method based on the alternating optimal solutions of all possible reusing pairs.

\subsection{MUD Matrix Optimization Subproblem}

In this subsection, the MUD Matrix optimization subproblem is researched for fixed $\mathbf{P}$, $\mathbf{\Phi}$ and $\mathbf{X}$.
To employ the SDR approach, ${\bf{\Gamma }}_i = {{\bf{f}}_i}{\bf{f}}_i^H \in\! {\mathbb{C}^{M\times M}}$ is defined, and subjects to ${{\bf{\Gamma }}_i} \succeq 0$ and $\textrm{Rank} ({\bf{\Gamma }}_i) = 1, \forall i$.
Moreover, ${\bf{e}}\! =\! {[{e_1}, \cdots ,{e_N}]^T}$ represents the vector consisting of the diagonal elements of matrix $\mathbf{\Phi}$ and ${\mathbf{G}_{i,r}}\! =\!\rm{ diag}({\mathbf{g}_{i,r}})$ represents the diagonal matrix of the vector ${\mathbf{g}_{i,r}}$.
Then, the channel gain from the $i$th CUE to gNB is reformulated as
\begin{align}
g_i^B &  = {\left| {{\bf{f}}_i^H\left( {{{\bf{g}}_{i,b}} + {\bf{g}}_{r,b}^H{\bf{\Phi }}{{\bf{g}}_{i,r}}} \right)} \right|^2} = {\left| {{\bf{f}}_i^H\left( {{{\bf{g}}_{i,b}} + {\bf{g}}_{r,b}^H{{\bf{G}}_{i,r}}{\bf{e}}} \right)} \right|^2}\nonumber\\
& = \left[ {1,{{\bf{e}}^H}} \right]{\left[ {{{\bf{g}}_{i,b}},{\bf{g}}_{r,b}^H{{\bf{G}}_{i,r}}} \right]^H}{{\bf{f}}_i}{\bf{f}}_i^H\left[ {{{\bf{g}}_{i,b}},{\bf{g}}_{r,b}^H{{\bf{G}}_{i,r}}} \right]{\left[ {1,{{\bf{e}}^H}} \right]^H}\nonumber\\
& = {\rm{Tr}} \left( \left[ {1,{{\bf{e}}^H}} \right]{{\bf{G}}_{i,B}}{\bf{\Gamma }}_i{{\bf{G}}_{i,B}}^H{\left[ {1,{{\bf{e}}^H}} \right]^H} \right)\nonumber\\
& = {\rm{Tr}}\left( {{\bf{\Gamma }}_i{\bf{G}}_{i,B}^H{\bf{\Lambda }_1}{{\bf{G}}_{i,B}}} \right) = {\rm{Tr}}\left( {{\bf{\Gamma }}_i{\bf{G}}_1} \right),
\end{align}
where ${{\bf{G}}_{i,B}}\! =\! {[{{\bf{g}}_{i,b}},{\bf{g}}_{r,b}^H{\bf{G}}_{j,r}]^H}\!\in\! {\mathbb{C}^{(N+1)\times M}}$,
${\bf{\Lambda }_1}={[1,{{\bf{e}}^H}]^H}[1,{{\bf{e}}^H}] \in\! {\mathbb{C}^{(N+1)\times (N+1)}}$ and  ${\bf{G}}_1={\bf{G}}_{i,B}^H{\bf{\Lambda }_1}{{\bf{G}}_{i,B}}$.
Similarly, the interference channel gain from the transmitter of the $l$th VUE pair to gNB is transformed into
\begin{align}
g_l^B &  = {\rm{Tr}}\left( {{\bf{\Gamma }}_i{\bf{G}}_{l,B}^H{\bf{\Lambda }_1}{{\bf{G}}_{l,B}}} \right) = {\rm{Tr}}\left( {\bf{\Gamma }}_i{{\bf{G}}_2} \right),
\end{align}
where ${{\bf{G}}_{l,B}}\! =\! {[{g_{l,b}},{\bf{g}}_{r,b}^H{{\bf{G}}_{l,r}}]^H} \in\! {\mathbb{C}^{(N+1)\times M}}$, ${\bf{G}}_2={\bf{G}}_{l,B}^H{\bf{\Lambda }_1}{{\bf{G}}_{l,B}}$ and ${{\bf{G}}_{l,r}}\! =\! \rm{ diag}({{\bf{g}}_{l,r}})$. Then, the subproblem is reformulated as
\begin{align}
\mathop {\max }\limits_{\{ {{\bf{\Gamma }}_i}\} } \quad  & W  {\log _2}\left( {1 + \frac{{P_i^c{\rm{Tr}}({{\bf{\Gamma }}_i}{{\bf{G}}_1})}}{{P_l^v{\rm{Tr}}({{\bf{\Gamma }}_i}{{\bf{G}}_2}) + {\sigma ^2}{\rm{Tr}}({{\bf{\Gamma }}_i})}}} \right)\label{fsdff}\\
\mathrm{s.t.}\quad  & {{\bf{\Gamma }}_i} \succeq 0,\tag{\ref{fsdff}{a}}\label{dsfgsda}\\
&\mathrm{Rank}({{\bf{\Gamma }}_i}) = 1.\tag{\ref{fsdff}{b}}\label{dsfgsda}
\end{align}
Problem (\ref{fsdff}) is a non-convex problem because the objective function and constraint (\ref{dsfgsda}) are non-convex.
The SCA method is employed to approximate the non-convex objective function with a convex function at each iteration.
Specifically, the objective function can be transformed into the form of the difference of two concave functions and the first order Taylor expansions of the concave functions at arbitrary points are globally upper bounded.
Therefore, the objective function at the given local point ${\bf{\Gamma }}_i^{(a)}$ in the current iteration number $a$ can be transformed into
\begin{align}
\!\!\!\!\!\!\!\!\!\!\!\!& W{\log _2}\left( {1 + \frac{{P_i^c{\rm{Tr}}\left( {{{\bf{\Gamma }}_i}{{\bf{G}}_1}} \right)}}{{P_l^v{\rm{Tr}}\left( {{{\bf{\Gamma }}_i}{{\bf{G}}_2}} \right) + {\sigma ^2}{\rm{Tr}}\left( {{{\bf{\Gamma }}_i}} \right)}}} \right)\quad\quad\quad\quad\quad\nonumber\\
 =  & W{\log _2}\left( {P_i^c{\rm{Tr}}\left( {{{\bf{\Gamma }}_i}{{\bf{G}}_1}} \right) + P_l^v{\rm{Tr}}\left( {{{\bf{\Gamma }}_i}{{\bf{G}}_2}} \right) + {\sigma ^2}{\rm{Tr}}\left( {{{\bf{\Gamma }}_i}} \right)} \right)\nonumber\\
 &- W{\log _2}\left( {P_l^v{\rm{Tr}}\left( {{{\bf{\Gamma }}_i}{{\bf{G}}_2}} \right) + {\sigma ^2}{\rm{Tr}}\left( {{{\bf{\Gamma }}_i}} \right)} \right)\nonumber\\
 \ge  & W{\log _2}\left( {P_i^c{\rm{Tr}}\left( {{{\bf{\Gamma }}_i}{{\bf{G}}_1}} \right) + P_l^v{\rm{Tr}}\left( {{{\bf{\Gamma }}_i}{{\bf{G}}_2}} \right) + {\sigma ^2}{\rm{Tr}}\left( {{{\bf{\Gamma }}_i}} \right)} \right)\nonumber\\
 &- W {{\log }_2}\left( {P_l^v{\rm{Tr}}\left( {{\bf{\Gamma }}_i^{(a)}{{\bf{G}}_2}} \right) + {\sigma ^2}{\rm{Tr}}\left( {{\bf{\Gamma }}_i^{(a)}} \right)} \right) \nonumber\\
 &- W\frac{{{\rm{Tr}}\left( {\left( {P_l^v{{\bf{G}}_2} + {\sigma ^2}{{\bf{I}}_M}} \right)\left( {{{\bf{\Gamma }}_i} - {\bf{\Gamma }}_i^{(a)}} \right)} \right)}}{{\left( {P_l^v{\rm{Tr}}\left( {{\bf{\Gamma }}_i^{(a)}{{\bf{G}}_2}} \right) + {\sigma ^2}{\rm{Tr}}\left( {{\bf{\Gamma }}_i^{(a)}} \right)} \right)\ln 2}} \nonumber\\
 =  & \tilde C_{i,l}^{a,1},\label{dggdgwegf}
\end{align}
which is a concave function with respect to the variable ${\bf{\Gamma }}_i$.
Then, the approximation of problem (\ref{fsdff}) is reformulated as
\begin{align}
\mathop {\max }\limits_{\{ {{\bf{\Gamma }}_i}\} } \quad  & \tilde C_{i,l}^{a,1}\label{dgdrhdhgf}\\
\mathrm{s.t.}\quad  & {{\bf{\Gamma }}_i} \succeq 0,\tag{\ref{dgdrhdhgf}{a}}\label{dssdfda}\\
&\mathrm{Rank}({{\bf{\Gamma }}_i}) = 1.\tag{\ref{dgdrhdhgf}{b}}\label{dgdfgvsda}
\end{align}
Furthermore, as the rank-one constraint in (\ref{dgdfgvsda}) is non-convex, we can temporarily relax it to write the relaxed problem of (\ref{dgdrhdhgf}) as
\begin{align}
\mathop {\max }\limits_{\{ {{\bf{\Gamma }}_i}\} } \quad  & \tilde C_{i,l}^{a,1} \label{bgdfgc}\\
{\rm{s}}{\rm{.t}}{\rm{.}}\quad  & {{\bf{\Gamma }}_i} \succeq 0.\tag{\ref{bgdfgc}{a}}\label{fdgdga}
\end{align}
Intuitively, problem (\ref{bgdfgc}) is a concave SDP problem and can be solved directly by CVX toolboxes.
When the rank of the optimal solution ${\bf{\Gamma }}_i^*$ to problem (\ref{bgdfgc}) satisfies $\mathrm{Rank}({\bf{\Gamma }}_i^*)=1$, the optimal solution ${\bf{\Gamma }}_i^*$ is also the optimal solution to problem (\ref{dgdrhdhgf}) and the optimal ${\bf{f}}_i^*$ can be obtained by applying the eigenvalue decomposition (EVD) of ${\bf{\Gamma }}_i^*$ \cite{chen2021qos}.
When $\mathrm{Rank}({\bf{\Gamma }}_i^*)>1$, problem (\ref{bgdfgc}) and problem (\ref{dgdrhdhgf}) are not necessarily equivalent.
Thus, we introduce a following lemma to elaborate that the rank-one constraint $\mathrm{Rank}({\bf{\Gamma }}_i^*)=1$ is satisfied in problem (\ref{bgdfgc}).
\begin{lemma}
The optimal solution to problem (\ref{bgdfgc}) is the optimal MUD matrix ${\bf{\Gamma }}_i^*$, which satisfies the rank-one constraint $\mathrm{Rank}({\bf{\Gamma }}_i^*)=1$.
\end{lemma}
\proof
Refer to Theorem 1 in \cite{chen2021qos}.
\endproof
Lemma 1 indicates that relaxing the rank-one constraint in problem (\ref{bgdfgc}) cannot result in a loss of the optimality compared to problem (\ref{dgdrhdhgf}).
In addition, based on the lower-bound adopted in (\ref{dggdgwegf}), any feasible solution to problem (\ref{dgdrhdhgf}) must be feasible for problem (\ref{fsdff}) as well.
Therefore, this means that the optimal solution to problem (\ref{dgdrhdhgf}) is a lower-bound of the optimal solution to problem (\ref{fsdff}).

\subsection{Phase Shift Optimization Subproblem}

In this subsection, the IRS phase shifts optimization subproblem is researched for given $\mathbf{P}$, $\mathbf{F}$ and $\mathbf{X}$.
To tackle the unit-modulus constraint in (\ref{dsfgsdb}), an auxiliary variable $u$ is introduced and the channel gain $g_i^B$ is reformulated as
\begin{align}
g_i^B &  = \left[ {u,{{\bf{e}}^H}} \right]{\left[ {{{\bf{g}}_{i,b}},{\bf{g}}_{r,b}^H{{\bf{G}}_{i,r}}} \right]^H}{{\bf{f}}_i}{\bf{f}}_i^H\left[ {{{\bf{g}}_{i,b}},{\bf{g}}_{r,b}^H{{\bf{G}}_{i,r}}} \right]{\left[ {u,{{\bf{e}}^H}} \right]^H}\nonumber\\
& = {\rm{Tr}}\left( {{{\bf{G}}_{i,B}}{\bf{\Gamma }}_i{\bf{G}}_{i,B}^H{\bf{\Lambda }}} \right) = {\rm{Tr}}\left( {{\bf{G}}_3{\bf{\Lambda }}} \right),
\end{align}
where ${\bf{\Lambda }}={[u,{{\bf{e}}^H}]^H}[u,{{\bf{e}}^H}] \in\! {\mathbb{C}^{(N+1)\times (N+1)}}$,  ${\bf{G}}_3={{\bf{G}}_{i,B}}{\bf{\Gamma }}_i{\bf{G}}_{i,B}^H$ and $u$ satisfies the constraint $u^2=1$.
Similarly, the interference channel gain $g_l^B$ is transformed into
\begin{align}
g_l^B &  = {\rm{Tr}}\left( {{{\bf{G}}_{l,B}}{\bf{\Gamma }}_i^H{\bf{G}}_{l,B}^H{\bf{\Lambda }}} \right) = {\rm{Tr}}\left( {{\bf{G}}_4{\bf{\Lambda }}} \right),
\end{align}
where ${\bf{G}}_4={{\bf{G}}_{l,B}}{\bf{\Gamma }}_i^H{\bf{G}}_{l,B}^H$.
Due to ${\bf{\Lambda }}={[u,{{\bf{e}}^H}]^H}[u,{{\bf{e}}^H}]$, ${\bf{\Lambda }}$ must satisfy the constraint ${\bf{\Lambda }}\succeq0$ and the rank-one constraint $\mathrm{Rank}({\bf{\Lambda }})=1$.
Moreover, the channel gain $g_l^v$ and $g_i^v$ are transformed into
\begin{align}
g_l^v &  = {\left| {{g_l} + {\bf{g}}_{r,l}^H{\bf{\Phi }}{g_{l,r}}} \right|^2} = {\left| {{g_l} + {\bf{g}}_{r,l}^H{{\bf{G}}_{l,r}}{\bf{e}}} \right|^2}\nonumber\\
& = \left[ {{g_l},{\bf{g}}_{r,l}^H{{\bf{G}}_{l,r}}} \right]{\left[ {u,{{\bf{e}}^H}} \right]^H}\left[ {u,{{\bf{e}}^H}} \right]{\left[ {{g_l},{\bf{g}}_{r,l}^H{{\bf{G}}_{l,r}}} \right]^H}\nonumber\\
& = {\bf{g}}_1^H{\bf{\Lambda }}{{\bf{g}}_1}
\end{align}
and
\begin{align}
g_i^v &={{\left| {{g_{i,l}} + {\bf{g}}_{r,l}^H{\bf{\Phi }}{{\bf{g}}_{i,r}}} \right|}^2}
= {{\left| {{g_{i,l}} + {\bf{g}}_{r,l}^H {{\bf{G}}_{i,r}}{\bf{e}}} \right|}^2} \nonumber\\
&  = \left[ {{g_{i,l}},{\bf{g}}_{r,l}^H{{\bf{G}}_{i,r}}} \right]{\left[ {u,{{\bf{e}}^H}} \right]^H}\left[ {u,{{\bf{e}}^H}} \right]{\left[ {{g_{i,l}},{\bf{g}}_{r,l}^H{{\bf{G}}_{i,r}}} \right]^H}\nonumber\\
& = {\bf{g}}_2^H{\bf{\Lambda }}{{\bf{g}}_2},
\end{align}
where ${{\bf{g}}_1} = {\left[ {{g_l},{\bf{g}}_{r,l}^H{{\bf{G}}_{l,r}}} \right]^H} \in\! {\mathbb{C}^{(N+1)\times 1}}$, ${{\bf{g}}_2} = {\left[ {{g_{i,l}},{\bf{g}}_{r,l}^H{{\bf{G}}_{i,r}}} \right]^H} \in\! {\mathbb{C}^{(N+1)\times 1}}$.
With the assistance of the auxiliary variable $u$, the unit-modulus constraint in (\ref{dsfgsdb}) can be transformed into a SOC constraint ${{{\bf{\Lambda }}_{\left[ {n,n} \right]}}}=1,\forall n$.
Then, the IRS phase shifts optimization subproblem can be equivalently written as
\begin{align}
\mathop {\max }\limits_{\{ {\bf{\Lambda }}\} } \quad  & W{\log _2}\left( {1 + \frac{{P_i^c{\rm{Tr}}\left( {{{\bf{G}}_3}{\bf{\Lambda }}} \right)}}{{P_l^v{\rm{Tr}}\left( {{{\bf{G}}_4}{\bf{\Lambda }}} \right) + {\sigma ^2}{\rm{Tr}}\left( {{{\bf{\Gamma }}_i}} \right)}}} \right)\label{hdgchrt}\\
{\rm{s}}{\rm{.t}}{\rm{.}}\quad  & \Pr \left\{ {\frac{{P_l^v{\bf{g}}_1^H{\bf{\Lambda }}{{\bf{g}}_1}}}{{P_i^c{\bf{g}}_2^H{\bf{\Lambda }}{{\bf{g}}_2} + {\sigma ^2}}}}\ge {\gamma _{th}} \right\} \ge 1 - \delta ,\tag{\ref{hdgchrt}{a}}\label{dgcdxwa} \\
&{{{\bf{\Lambda }}_{\left[ {n,n} \right]}}}  = 1,\quad n = 1, \cdots ,N + 1,\tag{\ref{hdgchrt}{b}}\label{dgcdxwbfsd}\\
&{\rm{Rank}}\left( {\bf{\Lambda }} \right) = 1,\tag{\ref{hdgchrt}{c}}\label{dgcdxwcfrfg}\\
&{\bf{\Lambda }}\succeq 0,\tag{\ref{hdgchrt}{d}}\label{dgcdxwdgfgd}
\end{align}
The challenge in problem (\ref{hdgchrt}) main lies in the chance constraint in (\ref{dgcdxwa}), the non-convex objective function and the rank-one constraint in (\ref{dgcdxwcfrfg}).
As can be seen from the previous subsection, the rank-one constraint can be relaxed in problem (\ref{hdgchrt}) and the objective function can be approximated as a concave function by the SCA approach.
Thus, similar to (\ref{dggdgwegf}), the objective function at a given feasible solution ${\bf{\Lambda }}^{(a)}$ in the current iteration number $a$ can be approximated as
\begin{align}
& W{\log _2}\left( {1 + \frac{{P_i^c{\rm{Tr}}\left( {{{\bf{G}}_3}{\bf{\Lambda }}} \right)}}{{P_l^v{\rm{Tr}}\left( {{{\bf{G}}_4}{\bf{\Lambda }}} \right) + {\sigma ^2}{\rm{Tr}}\left( {{{\bf{\Gamma }}_i}} \right)}}} \right)\nonumber\\
 =  & W{\log _2}\left( {\frac{{P_l^v{\rm{Tr}}\left( {{{\bf{G}}_4}{\bf{\Lambda }}} \right) + {\sigma ^2}{\rm{Tr}}\left( {{{\bf{\Gamma }}_i}} \right) + P_i^c{\rm{Tr}}\left( {{{\bf{G}}_3}{\bf{\Lambda }}} \right)}}{{P_l^v{\rm{Tr}}\left( {{{\bf{G}}_4}{\bf{\Lambda }}} \right) + {\sigma ^2}{\rm{Tr}}\left( {{{\bf{\Gamma }}_i}} \right)}}} \right)\nonumber\\
 =  & W{\log _2}\left( {P_l^v{\rm{Tr}}\left( {{{\bf{G}}_4}{\bf{\Lambda }}} \right) + {\sigma ^2}{\rm{Tr}}\left( {{{\bf{\Gamma }}_i}} \right) + P_i^c{\rm{Tr}}\left( {{{\bf{G}}_3}{\bf{\Lambda }}} \right)} \right)\nonumber\\
 &  - W{\log _2}\left( {P_l^v{\rm{Tr}}\left( {{{\bf{G}}_4}{\bf{\Lambda }}} \right) + {\sigma ^2}{\rm{Tr}}\left( {{{\bf{\Gamma }}_i}} \right)} \right)\nonumber\\
 \ge  & W{\log _2}\left( {P_l^v{\rm{Tr}}\left( {{{\bf{G}}_4}{\bf{\Lambda }}} \right) + {\sigma ^2}{\rm{Tr}}\left( {{{\bf{\Gamma }}_i}} \right) + P_i^c{\rm{Tr}}\left( {{{\bf{G}}_3}{\bf{\Lambda }}} \right)} \right)\nonumber\\
 &  - W{\log _2}\left( {P_l^v{\rm{Tr}}\left( {{{\bf{G}}_4}{{\bf{\Lambda }}^{(a)}}} \right) + {\sigma ^2}{\rm{Tr}}\left( {{{\bf{\Gamma }}_i}} \right)} \right)\nonumber\\
& - W\frac{{P_l^v{\rm{Tr}}\left( {{{\bf{G}}_4}\left( {{\bf{\Lambda }} - {{\bf{\Lambda }}^{(a)}}} \right)} \right)}}{{\left( {P_l^v{\rm{Tr}}\left( {{{\bf{G}}_4}{{\bf{\Lambda }}^{(a)}}} \right) + {\sigma ^2}{\rm{Tr}}\left( {{{\bf{\Gamma }}_i}} \right)} \right)\ln 2}}\nonumber\\
 =  & \tilde C_{i,l}^{a,2}. \label{dghjtyrqgd}
\end{align}
Because the chance constraint can be conservatively approaximated by the CVaR when the constraint function is either concave or quadratic with the random variables.
Then, the chance constraint in (\ref{dgcdxwa}) can be approximated by employing the CVaR method, which is given in the following lemma.
\begin{lemma}
Let $f(\mathbf{Z}) \in \mathbb{C}$, ${\bf{f}}(\mathbf{\mathbf{Z}})\! \in\! \mathbb{C}^{2(N+1)\times 1}$ and ${\bf{F}}(\mathbf{Z})\!\in \!\mathbb{H}^{2(N+1)}$ be the function with the variable matrix $\mathbf{Z} \in\mathbb{C}^{(N+1)}$, where $\mathbb{H}^{2(N+1)}$ is a $2(N+1)$-order hermitian matric. Suppose there exists a chance constraint formulated as
\begin{align}
\Pr \{ f(\mathbf{Z}) + {\bf{f}}(\mathbf{Z})^H{\bf{g}} + {{\bf{g}}^H}{\bf{F}}(\mathbf{Z}){\bf{g}} \le 0\}  \ge 1 - \delta, \label{ghfhrrtrewe}
\end{align}
where ${\bf{g}}\in \mathbb{C}^{2(N+1)\times 1}$ is a random vector. Then, constraint (\ref{ghfhrrtrewe}) can be approximated as
\begin{small}
\begin{equation}
\left\{
\begin{aligned}
&q + \frac{1}{\delta} \text{Tr} \left( {\boldsymbol{\Xi} \boldsymbol{\Psi}} \right) \le 0, \ q \in \mathbb{R}, \ \boldsymbol{\Psi} \in \mathbb{H}^{2N+3}, \ \boldsymbol{\Psi} \succeq 0,\\
&\boldsymbol{\Psi} - \left[ \begin{array}{cc}
\mathbf{F}(\mathbf{Z}) & \frac{1}{2} \mathbf{f}(\mathbf{Z}) \\
\frac{1}{2} \mathbf{f}(\mathbf{Z})^H & f(\mathbf{Z}) - q
\end{array} \right] \succeq 0,
\end{aligned}
\right.
\end{equation}
\end{small}
where $q$ is an auxiliary variable, ${\bf{\Psi}}$ is a matrix of auxiliary variables and ${\bf{\Xi}}$ is the second-order moment matrix of ${\bf{g}}$.
\end{lemma}
\proof
Refer to Theorem 1 of Section II in \cite{zymler2013distributionally}.
\endproof
In order to meet the conditions of application of Lemma 2, the function in the chance constraint can be transformed into
\begin{align}
 & \frac{{P_l^v{\bf{g}}_1^H{\bf{\Lambda }}{{\bf{g}}_1}}}{{P_i^c{\bf{g}}_2^H{\bf{\Lambda }}{{\bf{g}}_2} + {\sigma ^2}}} \ge {\gamma _{th}}\nonumber\\
 \Leftrightarrow  & {\gamma _{th}}\left( {P_i^c{\bf{g}}_2^H{\bf{\Lambda }}{{\bf{g}}_2} + {\sigma ^2}} \right) - P_l^v{\bf{g}}_1^H{\bf{\Lambda }}{{\bf{g}}_1} \le 0\nonumber\\
 \Leftrightarrow  & \left[\!\! {\begin{array}{*{20}{c}}
{{\bf{g}}_2^H}&{{\bf{g}}_1^H}
\end{array}} \!\!\right]\underbrace {\left[ {\begin{array}{*{20}{c}}
{{\gamma _{th}}P_i^c{\bf{\Lambda }}}&{\bf{0}}\\
{\bf{0}}&{ - P_l^v{\bf{\Lambda }}}
\end{array}} \right]}_{\bf{\Delta }\in {\mathbb{H}^{2(N+1)}}}\underbrace {\left[ {\begin{array}{*{20}{c}}
{{{\bf{g}}_2}}\\
{{{\bf{g}}_1}}
\end{array}} \right]}_{{{\bf{g}}_3}\in {\mathbb{C}^{2(N+1)\times1}}} \!\!+ {\gamma _{th}}{\sigma ^2} \le 0\nonumber\\
 \Leftrightarrow  & {\bf{g}}_3^H{\bf{\Delta }}{{\bf{g}}_3} + {\gamma _{th}}{\sigma ^2} \le 0,
\end{align}
which becomes a quadratic expression with the uncertain vector ${\bf{g}}_3$.
To acquire the second-order moment matrix of the uncertain channel gain ${\bf{g}}_3$, we need learn its probability distribution information.
More specifically, $S$ i.i.d. samples of the uncertain CSI ${\bf{g}}_3$ are collected and denoted as ${{\cal S}_v} = \left\{ {{\boldsymbol{\varsigma }}_1^v,{\boldsymbol{\varsigma }}_2^v, \cdots ,{\boldsymbol{\varsigma }}_S^v} \right\}$, where ${\boldsymbol{\varsigma }}_s^v \in {\mathbb{C}^{2(N + 1) \times 1}}$.
Then, we can obtain the mean vector and the covariance matrix of ${\bf{g}}_3$, i.e. ${{{\boldsymbol{\bar \varsigma }}}_v} = \frac{1}{S}\sum\limits_{s = 1}^S {{\boldsymbol{\varsigma }}_s^v} $ and ${\mathbf{\Sigma} _v} = \frac{1}{S}\sum\limits_{s = 1}^S {\left( {{\boldsymbol{\varsigma }}_s^v - {{{\boldsymbol{\bar \varsigma }}}_v}} \right){{\left( {{\boldsymbol{\varsigma }}_s^v - {{{\boldsymbol{\bar \varsigma }}}_v}} \right)}^H}}$, respectively.
Thus, the second-order moment matrix of  ${\bf{g}}_3$ can be expressed as
\begin{align}
{{\bf{\Xi }}_v} = \left[ {\begin{array}{*{20}{c}}
{{{\bf{\Sigma }}_v} + {{{\boldsymbol{\bar \varsigma }}}_v}{\boldsymbol{\bar \varsigma }}_v^H}&{{{{\boldsymbol{\bar \varsigma }}}_v}}\\
{{\boldsymbol{\bar \varsigma }}_v^H}&1
\end{array}} \right] \in {\mathbb{H}^{2N + 3}}.
\end{align}
From the above analysis, the chance constraint in (\ref{dgcdxwa}) is finally approximated as
\begin{small}
\begin{equation}
\!\!\left\{
\begin{aligned}
\nonumber
&q  + \frac{1}{\delta }{\rm{Tr}}\left( {{\bf{\Xi}_v \bf{\Psi}}} \right) \le 0,q  \in \mathbb{R},{\bf{\Psi}} \in {{\mathbb H}^{2N+3}},{\bf{\Psi}}\succeq0,\\
&{\bf{\Psi}} - \left[ {\begin{array}{*{20}{c}}
{{\bf{\Delta }}}&{\bf{0}}\\
{\bf{0}}&{{\gamma _{th}}{\sigma ^2} - q }
\end{array}} \right]\succeq0,
\end{aligned}
\right.
\end{equation}
\end{small}
Based on the above approximation and relaxation, problem (\ref{hdgchrt}) is reformulated as
\begin{align}
\mathop {\max }\limits_{\{ q,{\bf{\Lambda }},{\bf{\Psi }}\} } \quad  & \tilde C_{i,l}^{a,2} \label{dfggcbcc}\\
{\rm{s}}{\rm{.t}}{\rm{.}}\quad  & q + \frac{1}{\delta }{\rm{Tr}}\left( {{{\bf{\Xi }}_v}{\bf{\Psi }}} \right) \le 0,\tag{\ref{dfggcbcc}{a}}\label{fgdgdv}\\
&{\bf{\Psi }} - {\bf{\Omega }} \succeq 0,{\bf{\Psi }} \succeq 0,\;{\bf{\Psi }} \in {{\mathbb H}^{2N + 3}},\tag{\ref{dfggcbcc}{b}}\label{fgdgsffds}\\
&{{\bf{\Lambda }}_{\left[ {n,n} \right]}} = 1,\quad n = 1, \cdots ,N + 1,\tag{\ref{dfggcbcc}{c}}\label{fdffxvgv}\\
&{\bf{\Lambda }}\succeq 0,\tag{\ref{dfggcbcc}{d}}\label{ffdvbbndv}
\end{align}
where ${\bf{\Omega }} = \left[ {\begin{array}{*{20}{c}}
{\bf{\Delta }}&{\bf{0}}\\
{\bf{0}}&{{\gamma _{th}}{\sigma ^2} - q}
\end{array}} \right]$.
It can be seen that problem (\ref{dfggcbcc}) is a convex SDP problem, which can be efficiently tackled by applying CVX toolboxes.
Similarly, we can conclude that the optimal solution ${\bf{\Lambda }}^*$ must satisfy $\mathrm{Rank}({\bf{\Lambda }}^*)=1$ based on Lemma 1.
Thus, the optimal ${\bf{e}}^*$ can be obtained by means of using the EVD method to decompose ${\bf{\Phi }}^*$.
If $u=1$, ${\bf{e}}^*$ is equivalent to the optimal solution to problem (\ref{dfggcbcc}).
If $u=-1$, ${\bf{e}}^*$ is equivalent to the inverse of the optimal solution to problem (\ref{dfggcbcc}).

\subsection{Power Allocation Subproblem}
In this subsection, we research the power allocation subproblem based on the given variables $\mathbf{\Phi}$, $\mathbf{X}$ and $\bf{F}$ as follows
\begin{align}
\mathop {\max }\limits_{\{ {\bf{P}}\} } \quad  & W  {\log _2}\left( {1 + \frac{{P_i^cg_i^B}}{{P_l^vg_l^B + {\sigma ^2 |{\bf{f}}_i^H|^2}}}} \right)\label{dfgdfgvbnnm}\\
s.t.\quad  & \Pr \{ \frac{{P_l^vg_l^v}}{{P_i^cg_i^v + {\sigma ^2}}} \ge {\gamma _{th}}\}  \ge 1 - \delta ,\tag{\ref{dfgdfgvbnnm}{a}}\label{dgfbfgha}\\
 & 0 \le P_i^c \le P_{\max }^c,\;0 \le P_l^v \le P_{\max }^v.\tag{\ref{dfgdfgvbnnm}{b}}\label{ddfdffgb}
\end{align}
Due to the non-convexity of the objective function of problem (\ref{dfgdfgvbnnm}) and the chance constraint in (\ref{dgfbfgha}), we propose a self-learning method to solve this non-convex subproblem.
To be specific, to transform the chance constraint, the uncertain CSI of V2V links is denoted as a high probability region (HPR) where the QoS can be enforced by every CSI value.
The HPR can be obtained by collecting multiple samples of the uncertain CSI of V2V links and learning the uncertain set.
For ensuring the outage probability of V2V links, the uncertain set must cover the samples with a $1-\delta$ probability.
If the power allocation solution is feasible under all CSI of the uncertain set, the chance constraint must be satisfied.
Inspired by this thought, the power allocation subproblem can be approximated as
\begin{align}
\mathop {\max }\limits_{\{ {\bf{P}}\} } \quad W & {\log _2}\left( {1 + \frac{{P_i^cg_i^B}}{{P_l^vg_l^B + {\sigma ^2 |{\bf{f}}_i^H|^2}}}} \right)\label{fsedfff}\\
s.t.\quad  & {\bf{p}}_l^v{\boldsymbol{\varphi }}_l^v \ge {\sigma ^2},\;{\boldsymbol{\varphi }}_l^v \in {\mathcal{G}}_l^v,\tag{\ref{fsedfff}{a}}\label{affsdfa}\\
 & 0 \le P_i^c \le P_{\max }^c,\;0 \le P_l^v \le P_{\max }^v,\tag{\ref{fsedfff}{a}}\label{afsfafb}
\end{align}
where $\mathbf{p}_l^v = [\frac{{P_l^v}}{{\gamma _{th}}}, - P_i^c]$, $\boldsymbol{\varphi} _l^v = {[g_l^v,g_i^v]^T}$ and ${\mathcal{G}}_l^v$ is the HPR which needs to be learned.
It is not difficult to understand that if ${\mathcal{G}}_l^v$ covers the samples of the uncertain CSI $\boldsymbol{\varphi} _l^v$ with $1-\delta$ confidence level, any feasible solution to problem (\ref{fsedfff}) must satisfy
\begin{align}
\Pr \{ \frac{{P_l^vg_l^v}}{{P_i^cg_i^v + {\sigma ^2}}} \ge {\gamma _{th}}\}  \ge \Pr \{ \boldsymbol{\varphi}_l^d \in {\mathcal{G}}_l^v \} \ge 1 - \delta.
\end{align}
From the above analysis, a key point of solving problem (\ref{fsedfff}) is selecting a appropriately shaped HPR and learning the uncertainty parameters from it.
Because the affine set is a simple geometric and has excellent convexity, the HPR can be parameterized as
\begin{align}
{\mathcal{G}}_l^v = \left\{ {\boldsymbol{\varphi }}_l^v | {\bf{p}}_l^{v,(a)}{\boldsymbol{\varphi }}_l^v \ge {\kappa _l}  \right\},
\end{align}
where ${\bf{p}}_l^{v,(a)}=[\frac{{P_l^{v,(a)}}}{{\gamma _{th}}}, - P_i^{c,(a)}]$ represents a given feasible solution in the current iteration number $a$ and ${\kappa _l}$ denotes the size of ${\mathcal{G}}_l^v$.
In order to learn the size of HPR, we collect $S$ i.i.d. samples of the uncertain CSI ${\boldsymbol{\varphi }}_l^v$ as ${\cal S} = \left\{ {{{\boldsymbol{\varsigma }}_1},{{\boldsymbol{\varsigma }}_2}, \cdots ,{{\boldsymbol{\varsigma }}_s}} \right\}$, where ${{\boldsymbol{\varsigma }}_s} \in {\mathbb{R}^2}$.
Let $t({{\boldsymbol{\varsigma }}_s}) = {\bf{p}}_l^{v,(a)}{{\boldsymbol{\varsigma }}_s}$ be the transformation map from random space ${\mathbb{R}^2}$ into ${\mathbb{R}}$.
The size ${\kappa _l}$ can be obtained by estimating $(1-\delta)$-quantile of the underlying distribution of $t({{\boldsymbol{\varsigma }}_s})$ based on the sample set ${\cal S}$.
In order to achieve this goal, the $(1-\delta)$-quantile $w_{1-\delta}$ can be introduced as
\begin{align}
\Pr \{ t({\boldsymbol{\varsigma }}_s ) \le {w_{1 - \delta }}\}  = 1 - \delta.
\end{align}
Then, we can obtain the function values $t({\boldsymbol{\varsigma }})$ of all samples as $t({{\boldsymbol{\varsigma }}_1}), \cdots ,t({{\boldsymbol{\varsigma }}_s})$.
By sorting these values in ascending order ${t^{(1)}}({{\boldsymbol{\varsigma }}}) \le  \cdots  \le {t^{(s)}}({{\boldsymbol{\varsigma }}})$, the $\lceil {(1 - \delta )S} \rceil $-th value is acquired as the upper-bound of $w_{1-\delta}$.
Therefore, we can consider the size ${\kappa _l}$ of ${\mathcal{G}}_l^v$ as
\begin{align}
{\kappa _l} = {t^{\left( {\left\lceil {(1 - \delta )S} \right\rceil } \right)}}({\boldsymbol{\varsigma }}).
\end{align}
Based on the learned affine model, the robust counterpart of V2V QoS constraint is reformulated as
\begin{align}
\mathop {\min }\limits_{{\boldsymbol{\varphi }}_l^v} \quad  & {\bf{p}}_l^v{\boldsymbol{\varphi }}_l^v \label{sdgsrger}\\
{\rm{s}}{\rm{.t}}{\rm{.}}\quad \; & {\bf{p}}_l^{v,(a)}{\boldsymbol{\varphi }}_l^v \ge {\kappa _l}.\nonumber
\end{align}
Because problem (\ref{sdgsrger}) is feasible, the optimal objective value is acquired by its dual problem, which is formulated as
\begin{align}
\mathop {\max }\limits_{z_l^v} \quad  & z_l^v{\kappa _l} \label{svdfvdgfg}\\
{\rm{s}}{\rm{.t}}{\rm{.}}\quad \; & z_l^v{\bf{p}}_l^{v,(a)} \le {\bf{p}}_l^v,z_l^v \ge 0.\nonumber
\end{align}
Based on the above analysis, the chance constraint in (\ref{dgfbfgha}) can be converted into a combination of linear constraints as follow
\begin{subequations}  %´óÀ¨ºÅ¶à±àºÅ
\begin{numcases}{}
z_l^v{\kappa _l} \ge {\sigma ^2}  \label{fdfkdfkdjjf}\\
z_l^v{\bf{p}}_l^{v,(a)} \le {\bf{p}}_l^v,z_l^v \ge 0\label{jfsodjgoi}
\end{numcases}
\end{subequations}  %´óÀ¨ºÅ¶à±àºÅ
By means of decomposing the vectors ${\bf{p}}_l^{v,(a)}$ and ${\bf{p}}_l^v$ in (\ref{jfsodjgoi}), (\ref{jfsodjgoi}) is decomposed into
\begin{align}
\left\{ \begin{array}{l}
z_l^vP_l^{v,(a)} \le P_l^v\\
z_l^vP_i^{c,(a)} \ge P_i^c
\end{array} \right.\label{jfdxfgcvbi}
\end{align}
Then, by combining constraints (\ref{ddfdffgb}), (\ref{fdfkdfkdjjf}) and (\ref{jfdxfgcvbi}), the power allocation subproblem (\ref{dfgdfgvbnnm}) can be reformulated as
\begin{align}
\mathop {\max }\limits_{\{ P_i^c,P_l^v,z_l^v\} } \quad  & W  {\log _2}\left( {1 + \frac{{P_i^cg_i^B}}{{P_l^vg_l^B + {\sigma ^2 |{\bf{f}}_i^H|^2}}}} \right)\label{fdgfgdgbf}\\
{\rm{s}}{\rm{.t}}{\rm{.}}\;\;\quad  & 0 \le P_i^c \le \min \left\{ {z_l^vP_i^{c,(a)},P_{\max }^c} \right\},\tag{\ref{fdgfgdgbf}{a}}\label{fhbvbnvv}\\
 & z_l^vP_l^{v,(a)} \le P_l^v \le P_{\max }^v,\tag{\ref{fdgfgdgbf}{b}}\label{dfgbbnjjvv}\\
&z_l^v{\kappa _l} \ge {\sigma ^2},z_l^v \ge 0.\tag{\ref{fdgfgdgbf}{c}}\label{fdgfhgjyyuvv}
\end{align}
Because the objective function in (\ref{fdgfgdgbf}) is non-convex, a graph-based approach is proposed to solve the non-convex optimization problem.
From the constraints in (\ref{fdgfgdgbf}), we construct the feasible region of problem (\ref{fdgfgdgbf}) in Fig. 2 and then the power allocation is divided into two cases to discuss by comparing the values of $z_l^vP_i^{c,(a)}$ and $P_{\max }^c$ in (\ref{fhbvbnvv}).
\begin{figure}
\begin{center}
\includegraphics[scale = 0.35]{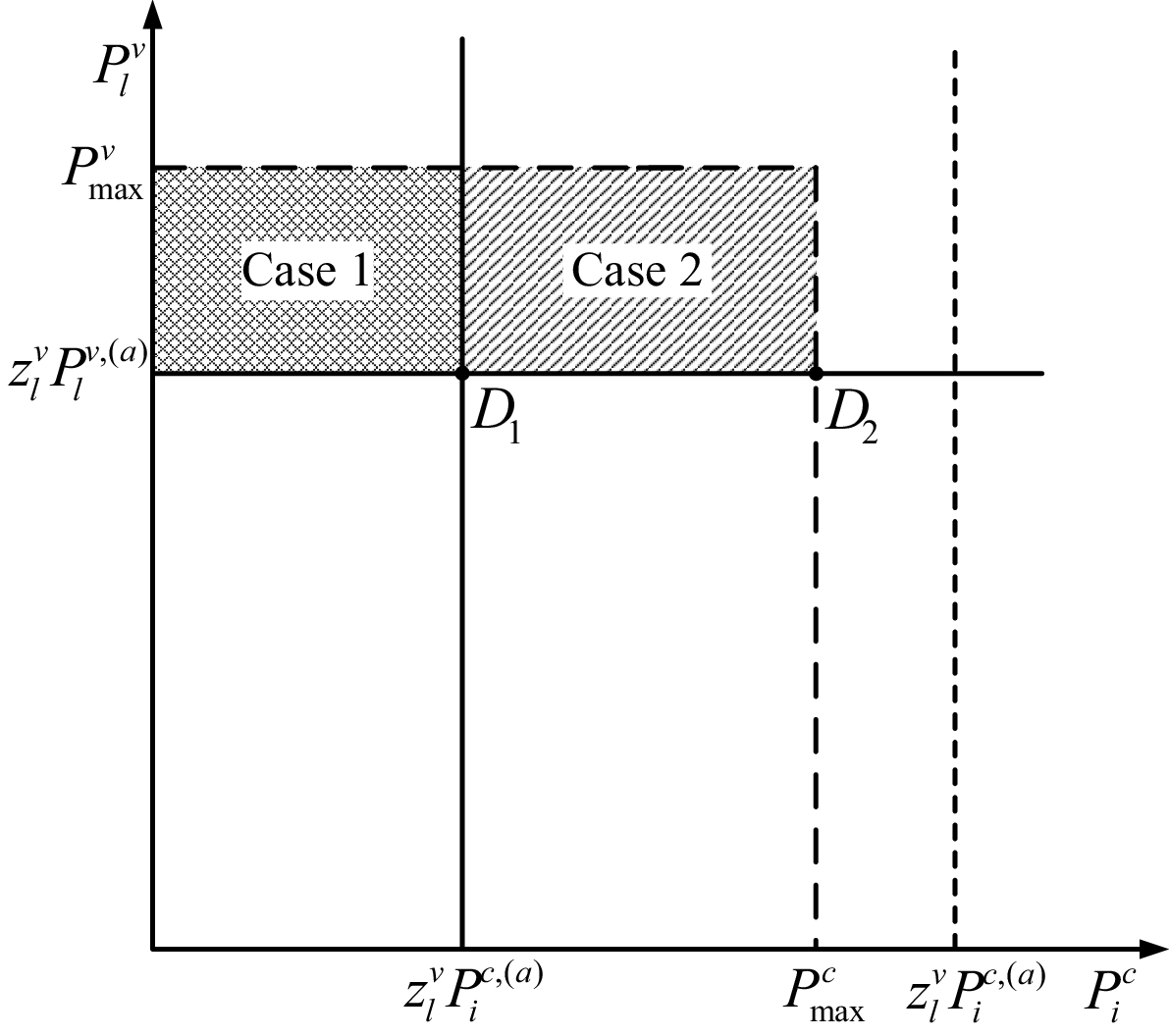}
\caption{Two cases in the feasible region of problem (\ref{fdgfgdgbf}).}\label{gdfsger}
\end{center}
%\vspace{-3em}
\end{figure}
\subsubsection{Case 1}
If $z_l^vP_i^{c,(a)} \le P_{\max }^c$, the feasible region satisfies Case 1 in Fig. 2.
By observing the objective function in (\ref{fdgfgdgbf}), it is not difficult to find that the objective function monotonically increases with $P_i^c$ when fixed $P_l^v$. In contrast, the objective function monotonically decreases with $P_l^v$ when fixed $P_i^c$.
Thus, we can conclude that the optimal solution in case 1 must be situated at ${D_1}$, which is the intersection point of $P_i^c=z_l^vP_i^{c,(a)}$ and $P_l^v=z_l^vP_l^{v,(a)}$.
Then, the coordinate of ${D_1}$ is substituted into the objective function to obtain a function of $z_l^v$, which is a monotonically increasing function.
Since the feasible region of $z_l^v$ in problem (\ref{fdgfgdgbf}) can be summarized as
\begin{align}
\frac{{{\sigma ^2}}}{{{\kappa_l}}} \le z_{\rm{l}}^v \le \min \left\{ {\frac{{P_{\max }^v}}{P_l^{v,(a)}},\frac{{P_{\max }^c}}{P_i^{c,(a)}}} \right\},
\end{align}
the optimal $z_l^{v,*}$ can be obtained as $z_l^{v,*}=\min \left\{ {\frac{{P_{\max }^v}}{P_l^{v,(a)}},\frac{{P_{\max }^c}}{P_i^{c,(a)}}} \right\}$.

\subsubsection{Case 2}
If $z_{\rm{l}}^v{P_i^{c,(a)}} \ge P_{\max }^c$, the feasible region satisfies Case 2 in Fig. 2.
Similar to the conclusions in case 1, the optimal solution in case 2 must be situated at $D_2$, which is the intersection point of $P_i^c=P_i^{max}$ and $P_l^v=z_l^vP_l^{v,(a)}$.
When the coordinate of $D_2$ is substituted into the objective function, the objective function becomes the function of variable $z_l^v$ and monotonically decreases with $z_l^v$.
Because the feasible region of $z_l^v$ in problem (\ref{fdgfgdgbf}) is summarized as
\begin{align}
\max \left\{ \frac{\sigma ^2} {\kappa_l},\frac{P_i^{max }} {P_i^{c,(a)}} \right\} \le z_{l}^v
\le \frac{P_l^{max }}{P_l^{v,(a)}},
\end{align}
the optimal $z_l^{v,*}$ can be obtained as $z_l^{v,*}=\max \left\{ {\frac{{{\sigma ^2}}}{{{\kappa_l}}},\frac{{P_{\max }^c}}{{P_i^{c,(a)}}}} \right\}$.

To sum up, by substituting $z_{l}^{v,*}$ into the coordinates of ${D_1}$ and ${D_2}$, the optimal power allocation $(P_i^{c,*},P_l^{v,*})$ of the subproblem is concluded as
\begin{small}
\begin{equation}
\!\!\!\!\!\!(P_i^{c,*},P_l^{v,*})\!\!=\!\!\left\{
\begin{aligned}
 \!\!(\frac{P_{\max }^v{P_i^{c,(a)}}}{{P_l^{v,(a)}}},P_{\max }^v) , &\textrm{if} \frac{\sigma ^2}{\kappa_l} \le \frac{P_{\max }^v}{P_l^{v,(a)}} \le \frac{{P_{\max }^c}}{{P_i^{c,(a)}}}, \\
 \!\!(P_{\max }^c,\frac{{P_{\max }^c}{P_l^{v,(a)}}}{P_i^{c,(a)}}) , &\textrm{if} \frac{{{\sigma ^2}}}{{{\kappa_l}}} \le \frac{{P_{\max }^c}}{{P_i^{c,(a)}}} \le \frac{{P_{\max }^v}}{P_l^{v,(a)}},\\
 \!\!(P_{\max }^c,\frac{{\sigma ^2}{P_l^{v,(a)}}}{{{\kappa_l}}}) ,\quad &\textrm{if} \frac{{P_{\max }^c}}{{P_i^{c,(a)}}} \le \frac{{{\sigma ^2}}}{{{\kappa_l}}} \le \frac{{P_{\max }^v}}{P_l^{v,(a)}}, \\
0, \quad\quad\quad\quad\quad\quad\quad&\textrm{otherwise}.  \label{fsfgvctrr}
\end{aligned}
\right.
\end{equation}
\end{small}

It is not difficult to find that computing the optimal power allocation solution from (\ref{fsfgvctrr}) only involves a few multiplication steps and no loop structures.
Thus, the computation complexity of the proposed power allocation approach is ${\cal O}(1)$.

\subsection{Spectrum Allocation Subproblem}

Through alternately iteratively optimizing the above three subproblems, we can obtain the maximum CUE capacity of all possible spectrum reusing pairs $\{{C_{i,l}^*}, \forall i,l\}$.
Then, the optimal spectrum reusing pattern can be acquired based on all maximum capacity of CUEs.
In practical V2X communication networks, the number of CUEs is usually greater than the number of VUE pairs.
This means that there are $I-L$ CUEs whose spectrum is not reused by any VUE pairs.
For formulating the spectrum allocation as a bipartite matching problem, the set ${\cal L}'$ containing the virtual VUE pairs is constructed as
\begin{small}
\begin{equation}
{{\cal L}'} = \!\left\{
\begin{aligned}
 & \{ L + 1, L + 2, \cdots, I\} & \text{if } I > L,\\
 & \phi & \text{if } I = L,
\end{aligned}
\right.
\end{equation}
\end{small}
where $\phi$ denotes a empty set.
The spectrum reusing pair between the $i$th CUE and the $l$th VUE pair ($l \in {{\cal L}'}$) means that the spectrum of the $i$th CUE is not shared with the $l$th VUE pair, thus the maximum transmit power $P_{max}^c$ can be used and the maximum capacity can be formulated as
\begin{align}
{C_{i,l}^*}=W{\log _2}\left( 1 + \frac{P_{\max }^cg_i^B}{\sigma ^2 |{\bf{f}}_i^H|^2} \right).
\end{align}
Then, the spectrum allocation subproblem can be written as
\begin{align}
\mathop {\max }\limits_{\bf{X}} \quad &\sum\limits_{i \in \cal I} \sum\limits_{l \in {L \cup {\cal L'}} } {{x_{i,l}}{C_{i,l}^*}}  \label{dfhnkdvnvjxclkv}\\
\quad \textrm{s.t.}\;\quad &\textrm{(\ref{1c})},\textrm{(\ref{1d})},\textrm{(\ref{fdfgre})},\nonumber
\end{align}
where ${C_{i,l}^*}$ denotes the obtained maximum capacity based on the given spectrum reusing pari $x_{i,l}$ by iteratively optimizing the variables $\mathbf{F}$, $\mathbf{\Phi}$ and $\mathbf{P}$.
This subproblem is a bipartite matching problem in graph, which can be solved by the Hungarian algorithm.
The computation complexity of the Hungarian algorithm is ${\cal O}(I^3)$.

\subsection{Convergence and Computational Complexity of ACAO Algorithm}

\begin{algorithm}[t]\label{Am1}
\caption{Adaptive Channel-Based Alternating Optimization (ACAO) Algorithm}
\begin{algorithmic}[1]
\STATE Initialize $\mathbf{F}^{(0)}$, ${{\bf{P}}^{(0)}}$ and ${{\bf{\Phi}}^{(0)}}$, and set the iterative index $a=0$.
\FOR {$i=1,\cdots,I$}
\FOR {$l=1,\cdots,L$}
\REPEAT
\STATE For given $\{{{\bf{P}}^{(a)}},{{\bf{F}}^{(a)}},{{\bf{\Phi}}^{(a)}}\}$, solve problem (\ref{bgdfgc}) to obtain the optimal solution ${{\bf{F}}^{(a+1)}}$;
\STATE For given $\{{{\bf{F}}^{(a+1)}},{{\bf{P}}^{(a)}},{{\bf{\Phi}}^{(a)}}\}$, solve problem (\ref{dfggcbcc}) to obtain the optimal solution ${{\bf{\Phi}}^{(a+1)}}$;
\STATE For given $\{{{\bf{F}}^{(a+1)}},{\bf{\Phi}}^{(a+1)},{{\bf{P}}^{(a)}}\}$,  obtain the optimal solution ${{\bf{P}}^{(a+1)}}$ from (\ref{fsfgvctrr});
\STATE $a=a+1$;
\UNTIL The variation of the objective values in two consecutive iterations is below a threshold $\varrho$;
\STATE The optimal capacity of the possible spectrum reusing pair $C_{i,l}^*$ is obtained;
\ENDFOR
\ENDFOR
\STATE Apply Hungarian algorithm to solve problem (\ref{dfhnkdvnvjxclkv}) and compute the optimal spectrum reusing pattern $\bf{X}^*$ based on $\{C_{i,l}^*,\forall i,l\}$.
\STATE Return the optimal solution $\{{{\bf{P}}^*},{{\bf{F}}^*},{{\bf{\Phi}}^*},\bf{X}^*\}$.
%\ENSURE The optimal resource allocation $\{\rho_{j,s}^*\}\{p_{j}^{c,*}\}\{p_{s}^{d,*}\}$.
\end{algorithmic}
\end{algorithm}
In this subsection, an alternating iterative optimization algorithm is proposed to solve problem (\ref{1}) and acquire a near-optimal solution.
For each possible spectrum pair, problem (\ref{dfsg}) is solved to obtain the maximum capacity $C_{i,l}^*$ by iteratively optimizing ${\mathbf{F},\mathbf{\Phi},\mathbf{P}}$ while fixed the other variables.
The Hungarian method is applied to obtain the optimal spectrum reusing pattern based on the all possible reusing pairs.
The detailed procedure of the proposed AO algorithm is summarized in \textbf{Algorithm 1}.

Then, the convergence of our proposed algorithm is analyzed as follows.
$C \left({{\bf{F}}^{(a)}},{{\bf{\Phi }}^{(a)}},{{\bf{P}}^{(a)}}\right)$ is defined as the objective function of problem (\ref{dfsg}) at the $a$-th iteration.
In line 5 of Algorithm 1, problem (\ref{bgdfgc}) is solved for given $\{{{\bf{P}}^{(a)}},{{\bf{\Phi}}^{(a)}}\}$ and the change of $C(\cdot)$ is written as
\begin{align}
 & C\left( {{{\bf{F}}^{(a)}},{{\bf{\Phi }}^{(a)}},{{\bf{P}}^{(a)}}} \right)\nonumber\\
\mathop  = \limits^{(u)}  & {C^1}\left( {{{\bf{F}}^{(a)}},{{\bf{\Phi }}^{(a)}},{{\bf{P}}^{(a)}}} \right)\nonumber\\
\mathop  \le \limits^{(v)}  & {C^1}\left( {{{\bf{F}}^{(a + 1)}},{{\bf{\Phi }}^{(a)}},{{\bf{P}}^{(a)}}} \right),
\end{align}
where ${C^1}\left( {{{\bf{F}}^{(a)}},{{\bf{\Phi }}^{(a)}},{{\bf{P}}^{(a)}}} \right)$ is the objective function of problem (\ref{bgdfgc}), $(u)$ issues from a theory that the first-order Taylor expansions are tight at a given point, $(v)$ holds because the optimal solution to problem (\ref{bgdfgc}) is obtained.
Likewise, in line 6, the variation of $C(\cdot)$ is written as
\begin{align}
 & C\left( {{{\bf{F}}^{(a + 1)}},{{\bf{\Phi }}^{(a)}},{{\bf{P}}^{(a)}}} \right)\nonumber\\
 =  & {C^2}\left( {{{\bf{F}}^{(a + 1)}},{{\bf{\Phi }}^{(a)}},{{\bf{P}}^{(a)}}} \right)\nonumber\\
 \le  & {C^2}\left( {{{\bf{F}}^{(a + 1)}},{{\bf{\Phi }}^{(a + 1)}},{{\bf{P}}^{(a)}}} \right),
\end{align}
where ${C^2}\left( {{{\bf{F}}^{(a + 1)}},{{\bf{\Phi }}^{(a)}},{{\bf{P}}^{(a)}}} \right)$ is the objective function of problem (\ref{dfggcbcc}).
In line 7, because problem (\ref{dfgdfgvbnnm}) can be solved optimally, the variation of $C(\cdot)$ is written as
\begin{align}
 & C\left( {{{\bf{F}}^{(a + 1)}},{{\bf{\Phi }}^{(a + 1)}},{{\bf{P}}^{(a)}}} \right)\nonumber\\
 \le  & C\left( {{{\bf{F}}^{(a + 1)}},{{\bf{\Phi }}^{(a + 1)}},{{\bf{P}}^{(a + 1)}}} \right).
\end{align}
To sum up, we can conclude as follows
\begin{align}
 & C\left( {{{\bf{F}}^{(a)}},{{\bf{\Phi }}^{(a)}},{{\bf{P}}^{(a)}}} \right)\nonumber\\
 \le  & C\left( {{{\bf{F}}^{(a + 1)}},{{\bf{\Phi }}^{(a + 1)}},{{\bf{P}}^{(a + 1)}}} \right),
\end{align}
which demonstrates that the objective function of problem (\ref{dfsg}) at each iteration is monotonically non-decreasing.
Because problem (\ref{dfhnkdvnvjxclkv}) is solved optimally be the Hungarian method and the transmit power of both CUE and VUE pair is limited, Algorithm 1 can eventually achieve convergence.

The computational complexity of \textbf{Algorithm 1} is then analysed below.
Because both problem (\ref{bgdfgc}) and problem (\ref{dfggcbcc}) involve linear matrix inequality (LMI) constraints that can be solved by a standard interior point method \cite{6891348}, a general expression to compute their computational complexity can be given as
\begin{align}
{\cal O}\!\!\left(\!\! {{{\left( {\sum\limits_{k = 1}^K {{\upsilon _k}}  } \right)}^{{1 \mathord{\left/
 {\vphantom {1 2}} \right.
 \kern-\nulldelimiterspace} 2}}}y\left( {{y^2} + y\sum\limits_{k = 1}^K {\upsilon _k^2}  +  \sum\limits_{k = 1}^K {\upsilon _k^3}   } \right)} \right)
\end{align}
where $y$ is the number of variables and $K$ is the number of LMIs of size $v_k$.
Specifically, since problem (\ref{bgdfgc}) contains one $M$-dimensional LMI and $M^2$ variables.
its complexity is ${{\cal O}_1}={\cal O}\left( {2{M^{5.5}} + {M^{4.5}}} \right)$.
Likewise, since problem (\ref{dfggcbcc}) contains $N+2$ $1$-dimensional LMIs, two $2N+3$-dimensional LMIs, one $N+1$-dimensional LMI and $1+(2N+3)^3+(N+1)^2$ variables,
its complexity of is ${{\cal O}_2}={\cal O}\left( {{{\left( {6N + 9} \right)}^{{1 \mathord{\left/
 {\vphantom {1 2}} \right.
 \kern-\nulldelimiterspace} 2}}}\eta \left( {{\eta ^2} \!+\! \eta \left( {{\eta _1} \!+\! 2\eta _2^2 + \eta _3^2} \right) \!+\! {\eta _1} \!+\! 2\eta _2^3 + \eta _3^3} \right)} \right)$,
where $\eta  = 1 + \eta _2^2 + \eta _3^2$, ${\eta _1} = N + 2$, ${\eta _2} = 2N + 3$ and ${\eta _3} = N + 1$.
Because the complexity of the power allocation subproblem is ${{\cal O}_3} = {\cal O}\left( 1 \right)$ and the complexity of the Hungarian algorithm is ${\cal O}(I^3)$, the complexity of the proposed AO algorithm is
${\cal O}\left( {{I^3} + IL{a_{\max }}\left( {{{\cal O}_1} + {{\cal O}_2}+{{\cal O}_3}} \right)} \right)$,
where ${{a_{\max }}}$ is the maximum number of iterations.

\section{Simulation Results}
\begin{figure}
\begin{center}
\includegraphics[scale = 0.3]{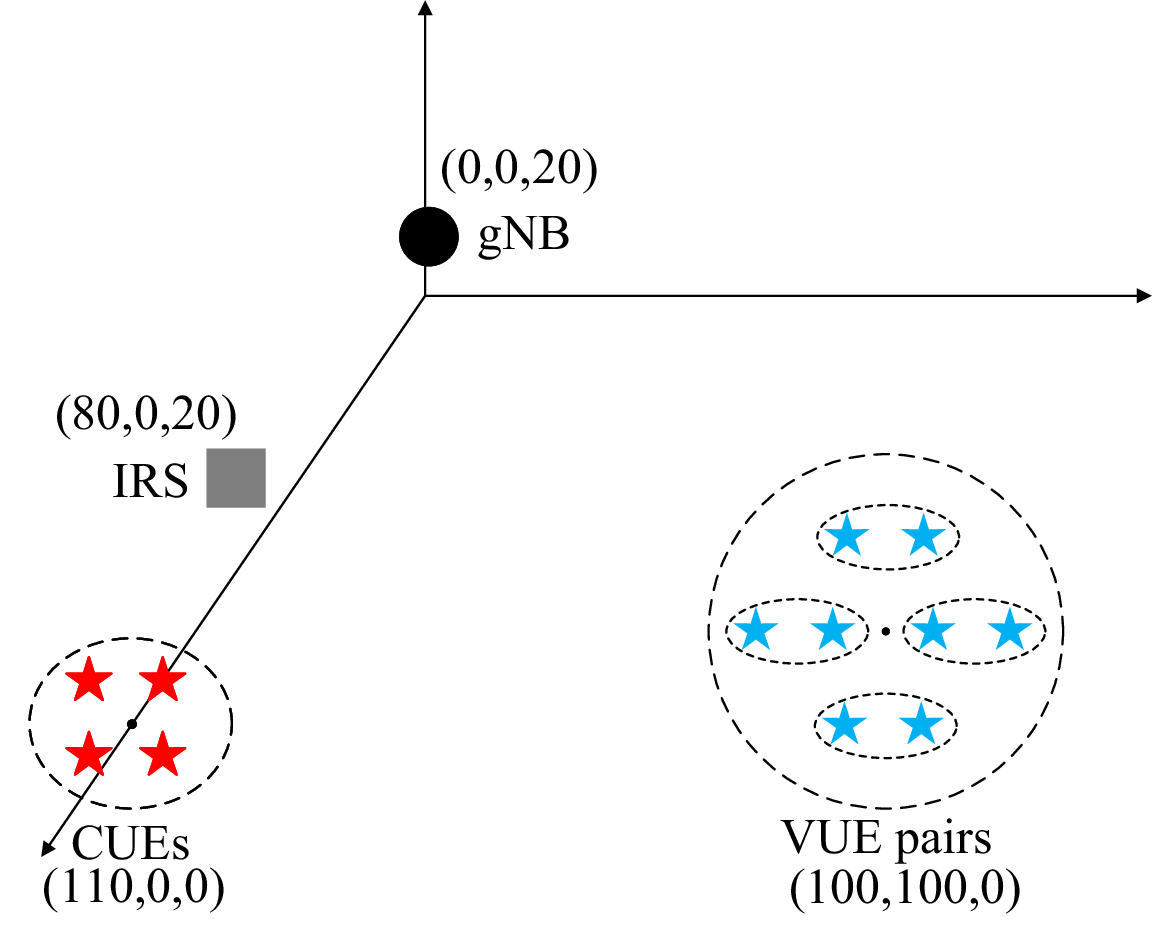}
\caption{Simulation model setup in IRS-aided V2X network.}\label{ffdgcvkh}
\end{center}
%\vspace{-2em}
\end{figure}
In this section, simulation results are provided to illustrate the performance of the proposed algorithm.
We model a IRS-aided vehicle network based on the freeway case of 3GPP TR 36.885 \cite{TSGRANV2X}, where the gNB model with approximately $250\mathrm{m}$ coverage is extensively applied in the simulation of vehicle networks.
In IRS-aided vehicle network, there are one gNB, one IRS, four CUEs and four VUE pairs as shown in Fig. \ref{ffdgcvkh}.
We assume that gNB is situated at $(0\mathrm{m},0\mathrm{m},20\mathrm{m})$ and IRS is situated at $(80\mathrm{m},0\mathrm{m},20\mathrm{m})$.
The gNB is equipped with $M=30$ antennas and the total number of IRS reflecting elements is $N=40$.
Moreover, the four CUEs are randomly generated in a circle centered at $(110\mathrm{m},0\mathrm{m},0\mathrm{m})$ with a radius of $40\mathrm{m}$ and the four VUE pairs are randomly generated in a circle centered at $(100\mathrm{m},100\mathrm{m},0\mathrm{m})$ with a radius of $60\mathrm{m}$.
The other important simulation parameters are given Table I.
Then, our proposed method is compared with the following four benchmark methods.
Specifically, the first benchmark method called Random Reuse is that VUE pairs randomly reuses the spectrum of CUEs, i.e., the spectrum allocation variables $\mathbf{X}$ are not optimized.
The second benchmark method called Random IRS is that the phase shift of the IRS reflecting elements is randomly assigned, i.e., the IRS phase shift matrix $\mathbf{\Phi}$ is not optimized.
The third benchmark method called Non-robust is that the small-scale fast fading effect is not considered.
The fourth benchmark method called No-IRS is that the vehicle network is not assisted by IRS.

\begin{table}[htb]
\caption{SIMULATION PARAMETERS}
\begin{tabular}{|c|c|}\hline
\textbf{Parameter}&\textbf{Value}\\\hline
        Bandwidth, $W$       &$10 \mathrm{MHz}$\\\hline
        Carrier frequency, $f_c$             &$2 \mathrm{GHz}$\\\hline
        Noise density, $\sigma^2$              &$-174 \mathrm{dBm}$\\\hline
        Period feedback delay, $T$        &$0.5 \mathrm{ms}$\\\hline
		Outage probability, $\delta $       &0.01 \\ \hline
        Vehicle speed, $\nu$                 &$80\ {\mathrm{km/h}}$\\ \hline
        SINR threshold, $\gamma _{th}$ &1\\ \hline
        %SINR threshold of V2V, $\Gamma _{\min }^d$ &1\\ \hline
        Maximum transmit power of CUE, $P_{\max }^c$  &$30\ {\mathrm{dBm}}$\\ \hline
        Maximum transmit power of VUE, $P_{\max }^v$  &$30\ {\mathrm{dBm}}$\\ \hline
        Pathloss exponent of gNB-vehicle             &4\\ \hline
        Pathloss exponent of IRS-vehicle             &2.2\\ \hline
        Pathloss exponent of gNB-IRS             &2\\ \hline
        Pathloss at the distance of 1m, $\alpha$   &$40\ {\mathrm{dB}}$\\ \hline
        %Shadowing standard deviation         &8 dB(CUE), 4 dB(VUE)\\ \hline
        Small-scale fast fading              &Rayleigh fading distribution\\ \hline
		Sample number, $S$                  &5000\\\hline
        %Sample number of the test set, $X$&6000\\\hline
        %Bisection search accuracy, $\xi$ &$10^{-4}$\\ \hline
        %Interior point method accuracy, $\iota$ &$10^{-3}$\\ \hline
        Convergence threshold, $\varrho$   &$10^{-3}$\\ \hline
\end{tabular}
%\vspace{-1em}
\end{table}

\begin{figure}
\begin{center}
\includegraphics[scale = 0.5]{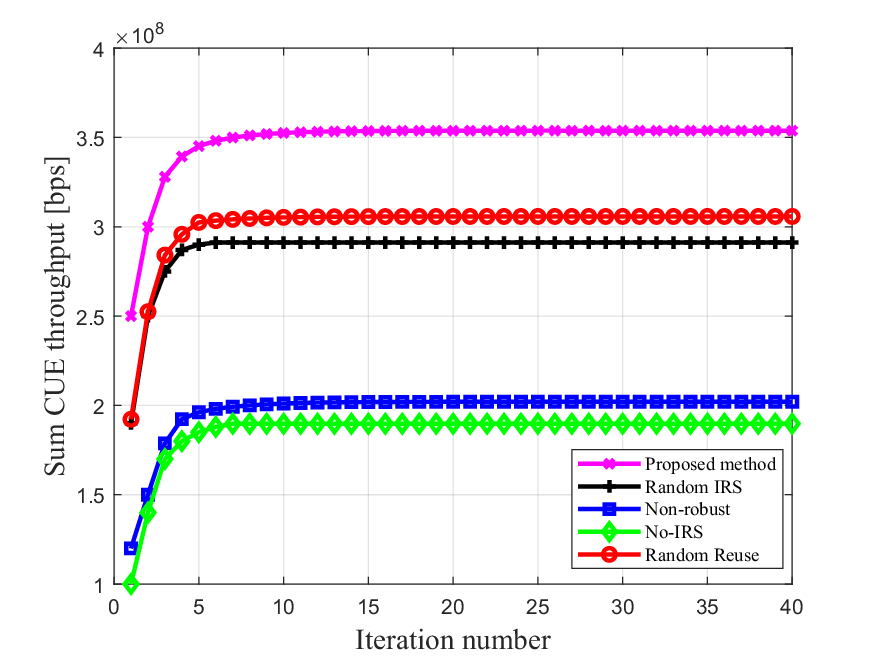}
\caption{The convergence of sum CUE throughput versus iteration number.}\label{sfdgcvrr}
\end{center}
%\vspace{-2em}
\end{figure}

Fig. \ref{sfdgcvrr} investigates the convergence of sum throughput of all CUEs versus the iteration number.
The convergence condition is as follows
\begin{align}
\sum\limits_{i = 1}^I {\left( {C_i^{(a + 1)} - C_i^{(a)}} \right) \leq \varrho } ,\nonumber
\end{align}
where ${C_i} = W{\log _2}(1 + {\gamma _i})$.
At the beginning of the iteration, the initial values of all variables are generated at random, so the sum  throughput of all methods is relatively small.
Then, with step-by-step iterations, the local optimal solution is acquired and the sum throughput of all methods gradually begins to converge to a stable value.
We can observe that all the curves are monotonically non-decreasing until the convergence condition is reached.
This phenomenon is consistent with the conclusions of the convergence analysis in Section V. E.
Moreover, it is not difficult to find that the performance of our proposed method is better than the performance of the other methods, and the performance of the methods with IRS-assistance are all better than that of the No-IRS method, which demonstrates that IRS can be applied to effectively improve the throughput of vehicular network.

\begin{figure}
\begin{center}
\includegraphics[scale = 0.5]{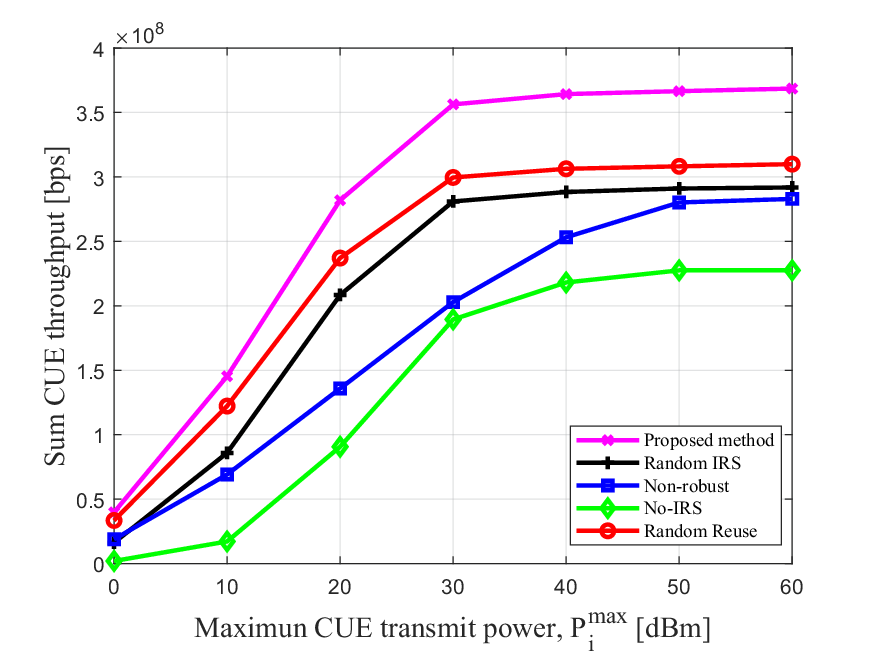}
\caption{Sum CUE throughput versus the maximum CUE transmit power.}\label{fkfvvhkh}
\end{center}
%\vspace{-2em}
\end{figure}
Fig. \ref{fkfvvhkh} shows the sum CUE throughput under variation of the maximum CUE transmit power. When the abscissa value is equal to $0\mathrm{dBm}$, we observe that the sum CUE throughput of all methods is quite small. Then, as the maximum transmit power increases, the sum CUE throughput of all methods begins to rise. When the abscissa value exceeds $40\mathrm{dBm}$, all the curves start to stabilize. The reason for this is that increased CUE transmit power can cause more interference to VUE pairs, which can result in the violation of the outage probability constraint in (\ref{1a}). To reduce the interference, the system cannot continue increasing the CUE transmit power, so the sum CUE throughput remains stable. It is not difficult to find that the throughput of the proposed method surpasses those of the Random IRS and Random Reuse methods, because their phase shift and spectrum allocation solutions are not optimal, respectively. Additionally, the throughput of the Non-robust method is lower than that of the other methods with IRS assistance since the outage probability constraint cannot be guaranteed by some samples of the uncertain channels.

\begin{figure}
\begin{center}
\includegraphics[scale = 0.5]{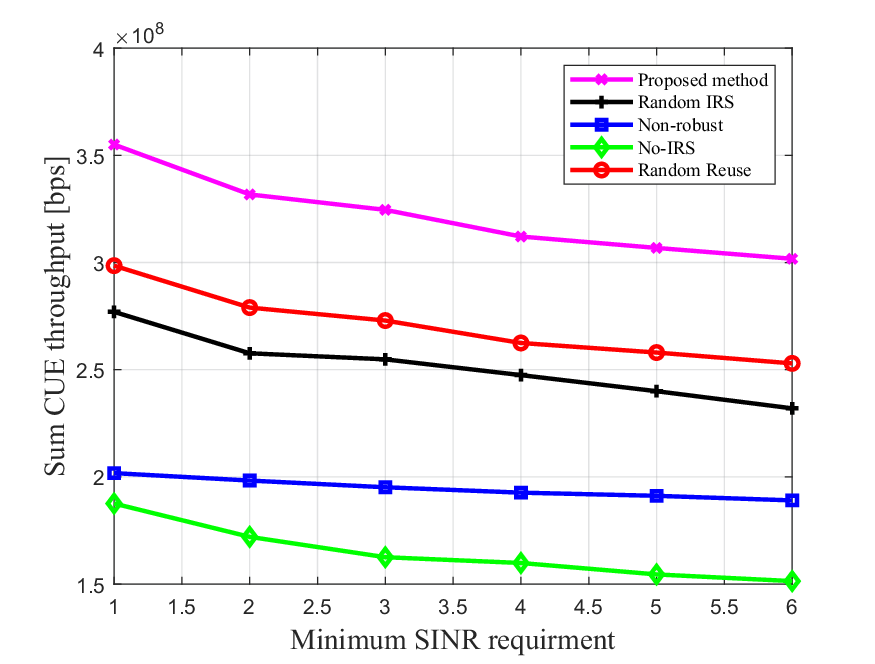}
\caption{Sum CUE throughput versus the minimum VUE SINR requirement.}\label{gdnghtd}
\end{center}
%\vspace{-2em}
\end{figure}
%Fig. \ref{gdnghtd} plots the impact of different minimum SINR requirement of VUE pairs on the sum CUE throughput.
%It is apparent that the sum CUE throughput of all methods decreases as the the minimum SINR threshold $\gamma_{th}$ increases.
%This is because the more VUE transmit power is allocated to satisfy the minimum SINR requirement of VUE pairs.
%The more VUE transmit power can result in the larger interference to CUEs that can make the the sum CUE throughput reduced.
%When the minimum SINR requirement increases from 1 to 6, the sum CUE throughput of the proposed method and No-IRS method decreases by $14.37\%$ and $19.62\%$, respectively.
%The $5.25\%$ gain difference is generated because of the performance gain of IRS-assistance vehicular communications.
Fig. \ref{gdnghtd} plots the impact of different minimum SINR requirements of VUE pairs on the sum CUE throughput. It is apparent that the sum CUE throughput of all methods decreases as the minimum SINR threshold $\gamma_{th}$ increases. This is because more VUE transmit power is allocated to meet the minimum SINR requirement of VUE pairs, resulting in larger interference to CUEs and reducing the sum CUE throughput. When the minimum SINR requirement increases from 1 to 6, the sum CUE throughput of the proposed method and No-IRS method decreases by $14.37\%$ and $19.62\%$, respectively. The $5.25\%$ gain difference is attributed to the performance gain of IRS-assisted vehicular communications.

\begin{figure}
\begin{center}
\includegraphics[scale = 0.5]{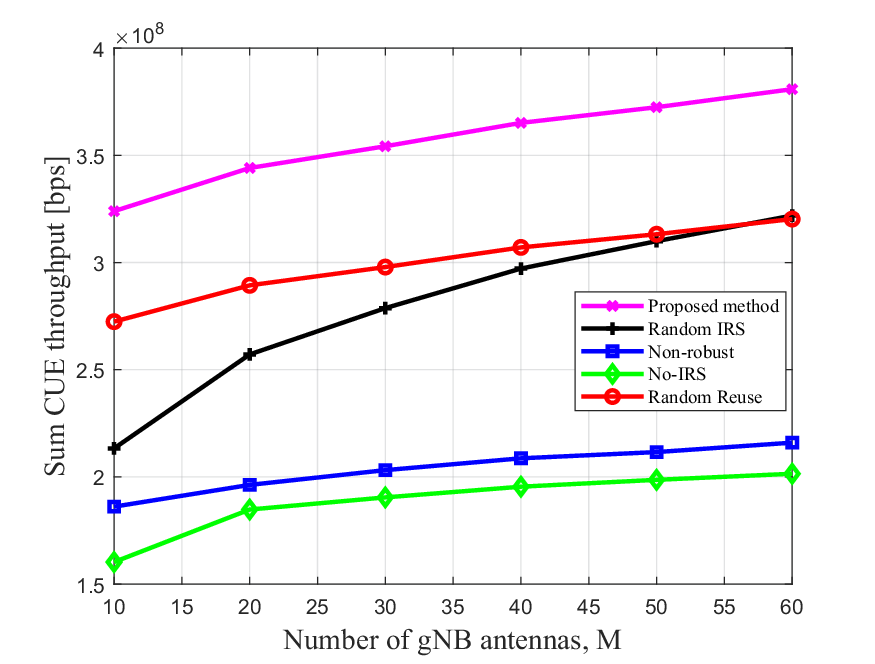}
\caption{Sum CUE throughput versus the number of gNB antennas $M$.}\label{dgfbvcbr}
\end{center}
%\vspace{-1.5em}
\end{figure}
%Fig. \ref{dgfbvcbr} illustrates the sum CUE throughput versus variation of the number of gNB antennas $M$.
%It is observed that the sum CUE throughput of all methods increases as the number of gNB antennas grows.
%The reason is that more gNB antennas can provide a higher diversity gain, which can bring larger CUE throughput.
%When the number of gNB antennas grows from 10 to 60, the sum CUE throughput of Random IRS method and No-IRS method increases by $46.43\%$ and $26.25\%$, respectively.
%The reason why the gain of Random IRS is larger than the gain of No-IRS method is that the existence of IRS can effectively improve the CUE throughput.
%In addition, the gain of CUE throughput of the proposed method $17.54\%$ is relatively small because the overall throughput of the proposed method has been significantly enhanced by the IRS phase shift optimization problem.
Fig. \ref{dgfbvcbr} illustrates the sum CUE throughput versus the variation in the number of gNB antennas $M$. It is observed that the sum CUE throughput for all methods increases as the number of gNB antennas grows. This increase is due to more gNB antennas providing higher diversity gain, which in turn boosts the CUE throughput. As the number of gNB antennas increases from 10 to 60, the sum CUE throughput of the Random IRS method and the No-IRS method increases by $46.43\%$ and $26.25\%$, respectively. The larger gain of the Random IRS method over the No-IRS method is attributed to the effectiveness of IRS in improving CUE throughput. Additionally, the gain of $17.54\%$ in CUE throughput for the proposed method is relatively smaller because the overall throughput has already been significantly enhanced through IRS phase shift optimization.

\begin{figure}
\begin{center}
\includegraphics[scale = 0.5]{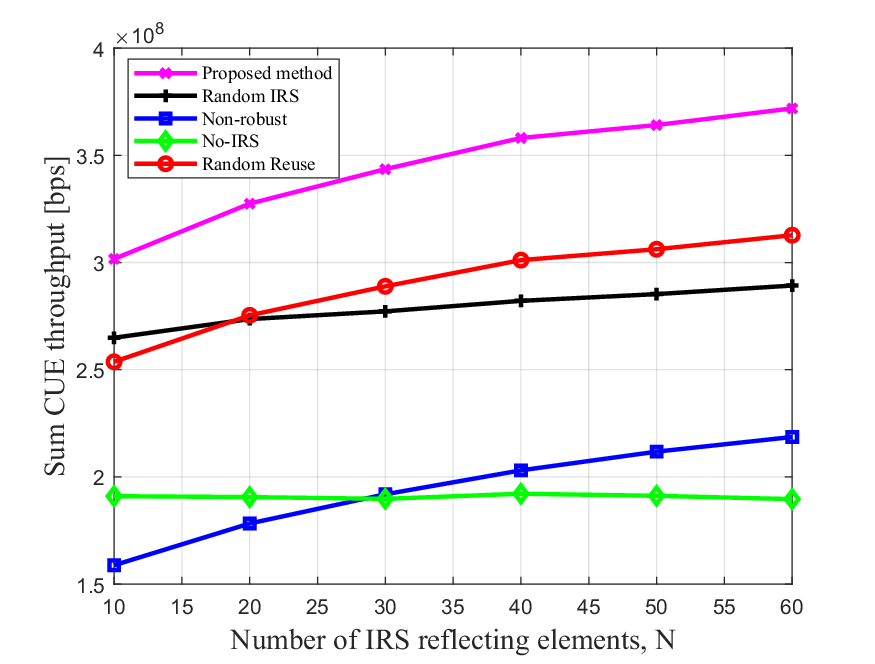}
\caption{Sum CUE throughput versus the number of IRS reflecting elements $N$.}\label{dfghjhggg}
\end{center}
%\vspace{-1.5em}
\end{figure}
%Fig. \ref{dfghjhggg} researches the sum CUE throughput versus variation of the number of IRS reflecting elements $N$.
%When the abscissa value increases from 10 to 60, the curves of all methods with IRS-assistance also gradually increases except the curve of No-IRS method.
%This is because a larger number of IRS reflecting elements can receive and reflect more signal energy, which can result in larger channel gain.
%The reason why the curve of No-IRS method remains constant is that IRS is not considered in the resource allocation problem of No-IRS method.
%It is observed that when the number of IRS reflecting elements increases from 10 to 60, the gain of CUE throughput of Random IRS method is smaller than the other methods with IRS-assistance.
%This is because the number of IRS reflecting elements has a relatively small effect on the gain of Random IRS when its IRS phase shift matrix $\mathbf{\Phi}$ is not optimized.
%Moreover, the gain of CUE throughput of the proposed method $22.7\%$ is significantly larger than the gain of the proposed method $17.54\%$ in Fig. \ref{dgfbvcbr}.
%This demonstrates that the gain obtained by increasing the number of reflecting elements $N$ is larger than the gain obtained by increasing the number of gNB antennas $M$.
%Because the cost of adding IRS reflecting elements to boost the same gain is much lower than the cost of adding the number of gNB antennas due to the low expense and difficulty of deploying an IRS, this simulation result suggests that IRS technology is superior to traditional massive MIMO in terms of energy efficiency.
Fig. \ref{dfghjhggg} investigates the sum CUE throughput versus the variation in the number of IRS reflecting elements $N$. When the abscissa value increases from 10 to 60, the curves of all methods with IRS assistance gradually increase, except for the curve of the No-IRS method. This is because a larger number of IRS reflecting elements can receive and reflect more signal energy, resulting in larger channel gains. The reason why the curve of the No-IRS method remains constant is that IRS is not considered in the resource allocation problem of the No-IRS method. It is observed that when the number of IRS reflecting elements increases from 10 to 60, the gain in CUE throughput of the Random IRS method is smaller than that of the other methods with IRS assistance. This is due to the relatively small effect of the number of IRS reflecting elements on the gain of the Random IRS when its IRS phase shift matrix $\mathbf{\Phi}$ is not optimized. Moreover, the gain in CUE throughput of the proposed method, $22.7\%$, is significantly larger than the gain of $17.54\%$ observed in Fig. \ref{dgfbvcbr}. This demonstrates that the gain obtained by increasing the number of reflecting elements $N$ is larger than the gain obtained by increasing the number of gNB antennas $M$. Given that the cost of adding IRS reflecting elements to achieve the same gain is much lower than the cost of adding gNB antennas due to the low expense and ease of deploying an IRS, this simulation result suggests that IRS technology is superior to traditional massive MIMO in terms of cost efficiency.

\begin{figure}
\begin{center}
\includegraphics[scale = 0.5]{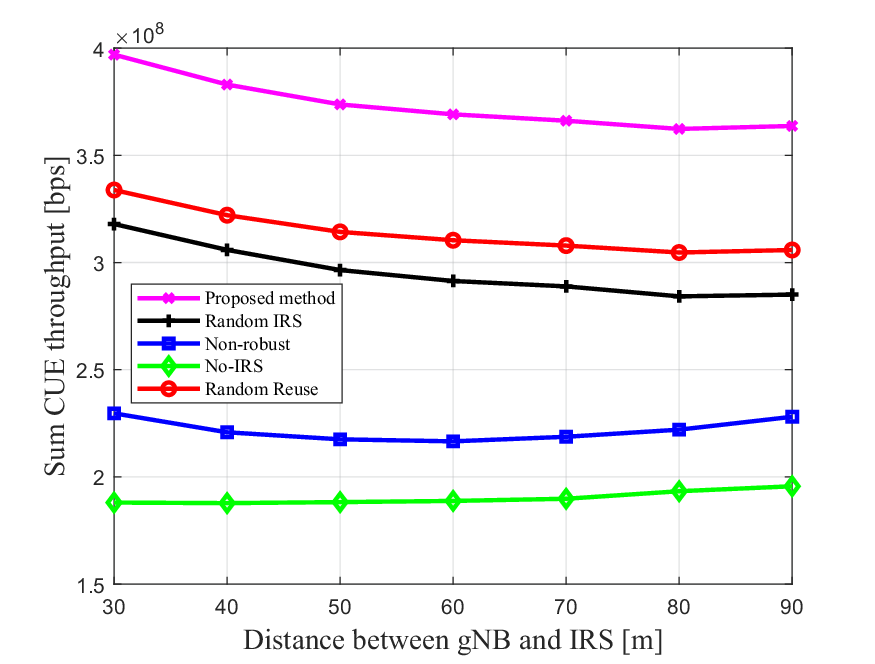}
\caption{Sum CUE throughput versus the distance between gNB and IRS.}\label{ggvcvbfgr}
\end{center}
%\vspace{-2em}
\end{figure}
In Fig. \ref{ggvcvbfgr}, we analyze the sum CUE throughput versus variation of the distance between gNB and IRS.
We assume that IRS is situated at $(x_{IRS},0\mathrm{m},20\mathrm{m})$, where $x_{IRS}$ denotes the distance between gNB and IRS.
From the parameter settings above, the distance between gNB and the centre of CUEs is approximately $d_{i,b}=120m$.
The pathloss exponents of gNB-IRS and IRS-vehicle are all set as $2$.
It can be seen that the sum CUE throughput of Non-robust method reduces form $30m$ to $60m$ and augments from $60m$ to $90m$.
Because the links gNB-IRS and IRS-CUE are cascaded channels, there exists the product pathloss \cite{wu2020joint}.
When gNB is located at the middle of the horizontal position of gNB and CUEs, the CUE capacity achieves the minimum. To reduce the product pathloss and increase the sum CUE capacity, IRS should be deployed close to gNB or CUEs.
We can also give the explanation through the following way,
This can also be explained through the following way, where the small-scale fading is neglected for the sake of simplicity.
The large-scale fading gain of the joint gNB-IRS-CUE channel can be reformulated as
\begin{align}
\sqrt {{\beta _{i,b}}}  = \sqrt {\alpha {{({d_{i,b}})}^4}}  + \sqrt {{\alpha ^2}{{({d_{i,r}})}^2}{{({d_{r,b}})}^{2}}} ,\label{gjdkgjsf}
\end{align}
where ${d_{i,b}} \approx {d_{i,r}} + {d_{r,b}}$.
Then, (\ref{gjdkgjsf}) can be regarded as the function of variable ${d_{r,b}}$.
By taking the derivative of variable ${d_{r,b}}$, we can find that when ${d_{i,r}} = {d_{r,b}}= \frac{1}{2}d_{i,b}$, the combined channel gain reaches a minimum value.
In addition, it can be seen that the sum CUE throughput of the proposed method, Random IRS method and Random Reuse method first decrease and then remain stable.
The reason is that these methods all consider the CSI uncertainty, which hinders the improvement of system  performance as the distance grows \cite{chen2021robust}.

\section{Conclusions}
In this work, we researched the robust resource allocation strategy for IRS-aided V2X communications, where the uncertainty of vehicular channels was considered due to the Doppler effect.
Based on the spectrum reusing model of CUEs and VUE pairs, we aimed to maximize the sum capacity of all CUEs, and guarantee the outage probability constraints of VUE pairs and the unit-modulus requirements of IRS.
Then, the resource allocation problem was decomposed into the subproblems of the transmit power allocation, the spectrum allocation, the MUD matrix and the IRS phase shifts optimization.
An alternating iterative optimization algorithm was developed to alternately optimize the four subproblems.
Simulation results showed that the proposed alternating optimization algorithm significantly enhanced the capacity of CUEs and was superior to other benchmark approaches.
In addition, the results also demonstrated that the performance of the vehicular network with IRS-assistance was much better than without IRS-assistance.
In the future, we intend to research the scenario where all channels are uncertain in IRS-aided V2X communications.

%\begin{appendices}
%      \section{Proof of Theorem 1}
%fsdfgsgsvsdfg
%\end{appendices}

\footnotesize
%\small
\bibliographystyle{IEEEtran}
\bibliography{references}

\end{document}